\documentclass[traditabstract]{aa}
\usepackage[colorlinks, linkcolor=black,  anchorcolor=black, citecolor=black]{hyperref}
\usepackage{subfigure}
\usepackage{graphicx}
\usepackage{txfonts}
\usepackage{longtable}
\usepackage{pdflscape}
\usepackage{hhline}
\usepackage{natbib}
\newcommand{\kms}{\hbox{km\,s$^{-1}$}}
\newcommand{\logg}{\hbox{log\,$\it g$}}
\newcommand{\feh}{\hbox{$\rm [M/H]$}}
\newcommand{\teff}{\hbox{$T_{\rm eff}$}}

\begin{document}

\title{Ultracool dwarfs identified using spectra in LAMOST DR7
\thanks{Tables 1 to 3 are only available in electronic form
at the CDS via anonymous ftp to cdsarc.u-strasbg.fr (130.79.128.5)
or via http://cdsweb.u-strasbg.fr/cgi-bin/qcat?J/A+A/}}%

\author{
You-Fen Wang\inst{1}
\and
A-Li Luo\inst{1,2 }\thanks{Corresponding Author: A-Li Luo  ~~  email: lal@nao.cas.cn}
\and
Wen-Ping Chen\inst{3,4}\thanks{Wen-Ping Chen ~~  email: wpchen@ncu.edu.tw}
\and
Hugh R. A. Jones\inst{5}
\and
Bing Du\inst{1}
\and
Yin-Bi Li\inst{1}
\and
Shuo Zhang\inst{6,7}
\and
Zhong-Rui Bai\inst{1}
\and
Xiao Kong\inst{1}
\and
Yan-Xin Guo\inst{1}
}

\institute{
CAS Key Laboratory of Optical Astronomy, National Astronomical Observatories, Beijing 100101, China.\\
\and
School of Astronomy and Space Science, University of Chinese Academy of Sciences, Beijing 100049, China\\
\and
Graduate Institute of Astronomy, National Central University, 300 Zhongda Road, Zhongli, Taoyuan 32001, Taiwan\\
\and
Department of Physics, National Central University,300 Zhongda Road, Zhongli, Taoyuan 32001, Taiwan\\
\and
School of Physics, Astronomy and Mathematics, University of Hertfordshire, College Lane, Hatfield AL10 9AB, UK\\
\and
Department of Astronomy, School of Physics, Peking University, Beijing 100871, China\\
\and
Kavli institute of Astronomy and Astrophysics, Peking University, Beijing 100871, China\\
}

\date{Received September 00, 0000; accepted March 00, 0000}

  \abstract
  {
In this work, we identify 734 ultracool dwarfs with a spectral type of M6 or later, including one L0. Of this sample, 625 were studied spectroscopically for the first time. All of these ultracool dwarfs are within 360~pc, with a \textit{Gaia} G magnitude brighter than $  \sim19.2$~mag. By studying the spectra and checking their stellar parameters (\teff, \logg,~ and \feh) derived with the LAMOST pipeline, we found their cool red nature and their metallicity to be consistent with the nature of Galactic thin-disk objects. Furthermore, 77 of them show lithium absorption lines at 6708~\AA,\  further indicating their young ages and substellar nature. Kinematics obtained through LAMOST radial velocities, along with the proper motion and parallax data from \textit{Gaia} EDR3, also suggest that the majority of our targets are thin-disk objects. Kinematic ages were estimated through the relationship between the velocity dispersion and the average age for a certain population. Moreover, we identified 35 binaries, with 6 of them reported as binaries for the first time.
  }

   \maketitle
\section{Introduction}

Ultracool dwarfs (hereafter UCDs) are usually defined as objects with a spectral type of M7 V or later \citep{1997AJ....113.1421K}, comprising very low-mass stars, brown dwarfs (BDs), and planetary-mass objects. The M7 V spectral type BDs are almost the highest-mass BDs that are just below the least mass stellar boundary. While the main-sequence stars, BDs, and planetary-mass objects are distinguished pedagogically by their masses, the mass values are not readily measurable. The luminosity and spectral type of a star, once core hydrogen fusion begins, remain effectively the same (depending on the mass) for hundreds of thousands to hundreds of million years on the main sequence. In contrast, for a substellar object, theoretical modeling indicates that -- to use an example from \citet{1997ApJ...491..856B} -- a maximum-mass BD (about 75~$M_J$) has a steady luminosity of $\log L/L_\odot   \sim -1.3$ to $\log L/L_\odot   \sim -2$ and an effective temperature of about $T_{\rm eff} \sim3000$~K for the first 10~Myr, then dropping to $\log L/L_\odot  \sim-4$ and $T_{\rm eff}  \sim 2200$~K by the time the object becomes 1~Gyr old, corresponding to a spectral-type change from about an M6 to an L0. This age dependence on the spectral type makes it difficult to identify relatively massive and hot substellar objects. Thus, the class of late M to mid-L dwarfs represents a mixture of low-mass stars and BDs. Objects with ages later than mid-L spectral type are certainly BDs.

Brown dwarfs are objects that are not massive enough to sustain core hydrogen fusion, namely, they are substellar objects with masses of $M \lesssim 75$--80$~M_J$ \citep{2021ApJ...920...85M}, where $M_J$ is the Jupiter mass. Theories indicate that the BDs with masses heavier than $  \sim 65 M_J$ \citep{1985ApJ...296..502D,2015A&A...577A..42B} have cores hot enough to fuse lithium -- and those above $  \sim 13M_J$ may ignite deuterium. Other than these short-lasting fusion reactions, a BD continues to cool and fades monotonically in terms of brightness. Those with $M \lesssim 13 M_J$ have no fusion reaction at all and are called ``planetary-mass objects'' (or free-floating planets) that are in contrast to exoplanets orbiting hosting stars. The first M-type BD, an M8, was discovered by \citet{1995Natur.377..129R} in the nearby cluster the Pleiades. In the same year, another BD of the T type (T6.5) was found by \citet{1995Natur.378..463N}, which is a companion to an M1 dwarf. These discoveries came 30 years after the theoretical prediction of BDs by \citet{1963ApJ...137.1121K}. Because of their intrinsic faintness, any optical spectroscopic vindication of a BD requires large-aperture telescopes equipped with sensitive instruments. In spite of thousands of ultracool dwarfs proposed thus far, the key to true substellar nature by way of spectroscopic confirmation is possible only among nearby targets and -- with dedicated efforts -- in nearby star-forming regions.

Here, we present ultracool dwarfs with spectra types mainly ranging from M6 V to M9V that have been found with the Large Sky Area Multi-Object Fiber Spectroscopic Telescope \citep[LAMOST,][]{2015RAA....15.1095L}. Thanks to a 4-m aperture and a 5-deg field of view to accommodate more than 4000 fibers on the focal plane, LAMOST is efficient in providing data for spectroscopic sky surveys.  In this work, we examined a set of more than 10,000 spectra for objects with late M spectral type from the LAMOST DR7\footnote{http://dr7.lamost.org} (hereafter LDR7) and excluded the giants, those with low signal-to-noise (S/N) spectra or bad spectra, and those with earlier spectral types than M6 V. This list of ultracool dwarfs, while serendipitous in its sky coverage that is set by the LAMOST survey footprints, have been spectroscopically identified, therefore, expanding their currently known inventory of the ultracool population in the solar neighborhood considering the small population of late M dwarfs comparing with the early ones. This sample represents the low mass stars and bright brown dwarfs, which is crucial to understanding the physical process at the stellar and substellar boundary. It is also very useful for exoplanet surveys since they are much fainter than solar-like stars and, hence, it is easier to detect the low-mass rocky planets orbiting around them.

This paper is organized as follows. In Sect.~\ref{sec:data}, we describe the spectral data set used in this work and the procedure to select the ultracool dwarfs. In Sect.~\ref{sec:photometry} we present the properties of the sample, such as the positions in the \textit{Gaia} color-absolute magnitude diagram (CAMD), to verify their cool and red atmospheres. In Sect.~\ref{sec:spectrascopic}, we diagnose the stellar parameters and compare them according to literature and we also inspect the possible existence of the lithium absorption line as an indication of the young age or substellar nature. In Sect.~\ref{sec:kinematic}, we perform kinematic analysis using the radial velocity (RV) provided by LDR7 and the five astrometric parameters from the \textit{Gaia} Early Data Release 3 \citep[hereafter GEDR3]{2021A&A...649A...1G} to infer their stellar population and kinematical age. In Sect.~\ref{sec:cpm}, we report 35 binaries, of which 14 are newly discovered. Finally, we summarize our work in Sect.~\ref{summary}.

\section{Data and sample selection}
\label{sec:data}

After the first light in 2008, commissioned in 2009--2010, followed by the Pilot Survey for one year \citep{2015RAA....15.1095L}, LAMOST had completed its five-year regular survey by June 2017. It is presently undertaking the second five-year regular survey. The latest data release, LDR7, includes low-resolution ($R  \sim ~1800$) spectra from which the stellar parameters were derived and which form the basis of the work reported here. With a practical LAMOST limiting sensitivity of  \textit{Gaia} $G \lesssim19.2$~mag, our ultracool dwarfs were randomly observed in the footprints of the LAMOST survey,  and the targets are randomly selected in any particular pointing within 5 square degrees. Information on the fiber assignment can be reviewed in \citep{2015RAA....15.1095L}.  In this paper, we describe in detail on how to identify ultracool M dwarfs in LDR7 using spectral features.

\subsection{LAMOST dataset}

In order to construct our ultracool dwarf sample, we decided to set the cut at M6 to account for spectral typing uncertainties introduced in the LAMOST analysis pipeline (1 - 5 subtype depending S/N ), as well as for spectral type dependence on age for BDs, so as not to exclude possible young BDs. In LDR7, 11,412 spectra were classified as being M6 or later by the LAMOST classification pipeline. Among these, about 33\% are classified as dwarfs (listed as ``dM'' in the official catalogue) and 67\% as giants (listed as ``gM'' in the official catalogue). In contrast, among early-type M spectra,  only about 4\% are found to be giants. The high fraction of giants at late-type M spectra type is attributed to the selection effect.

We cross-matched each of the 11,412 entries with the GEDR3 with a 5\arcsec\ matching radius within the LAMOST coordinates (J2000). Firstly,  469 targets that pipeline classified as late-type M and have Gaia color BP-RP$<1$ were removed because of a misclassification due to low spectra quality or data reduction. Secondly, we visually inspected all the spectra and winnowed out giants and early-type dwarfs via their spectral features, and we also excluded those having low signal-to-noise ratios. The remaining 734 targets are checked one by one to ascertain photometric counterparts to avoid mismatch caused by the multi-to-one problem, judged both by the positions and the colors of the 2MASS and the LAMOST targets.  We further discuss  the sample selection and present some details below.

\subsection{Spectral features of giants versus dwarfs}
\label{sec:feature}

\begin{figure}
\centering
\includegraphics[width=89.3mm]{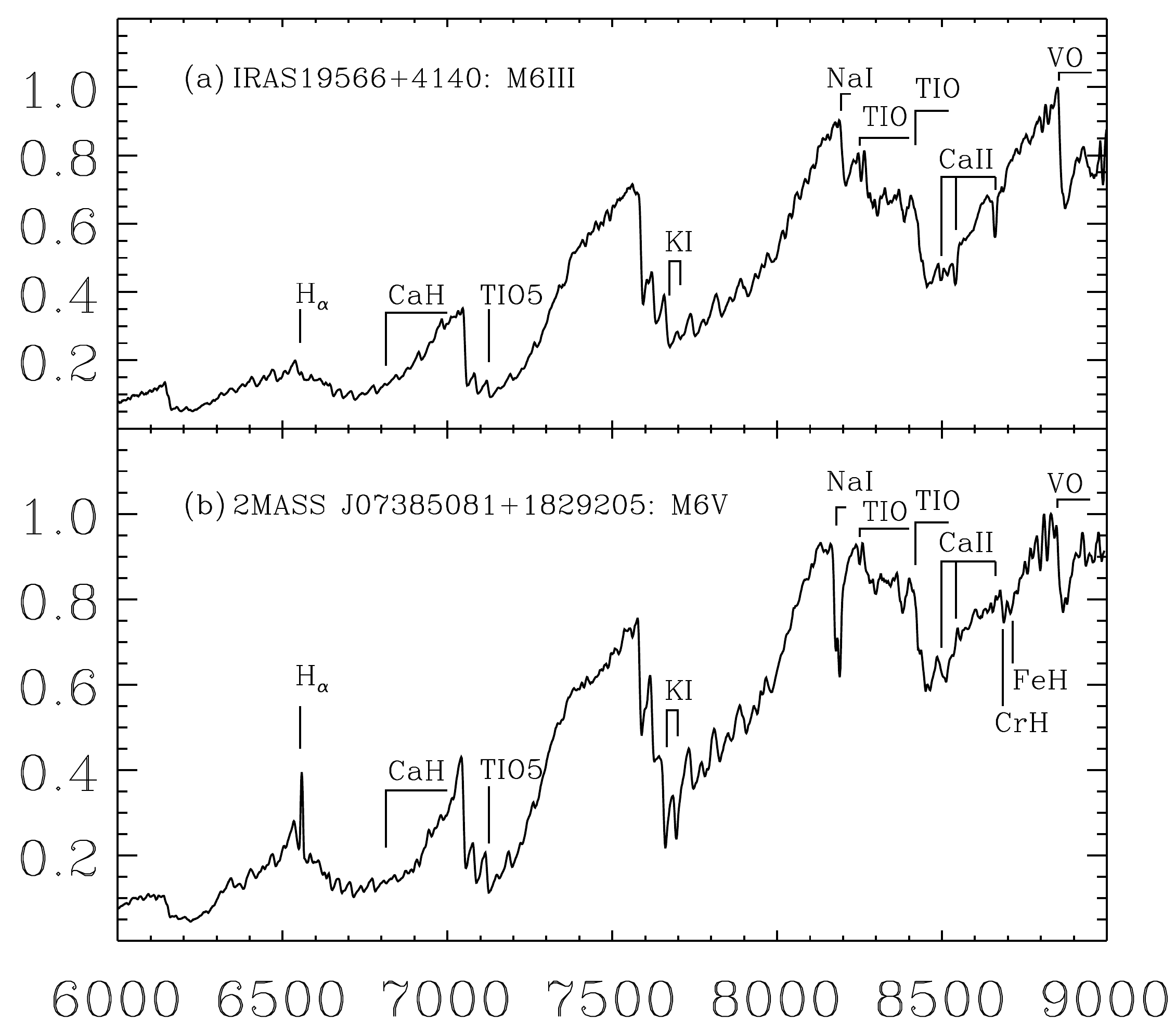}
\caption{ \label{fig:dwarf_vs_giant} Typical LAMOST spectrum of (a)~an M6\,III giant \citep{2016A&A...594A..39F}, and (b)~an M6 dwarf \citep{2014AJ....147...20N}. Both spectra have S/N greater than 100 in the $z$  band. }
\end{figure}

A general description of the spectral features of M types in the LAMOST spectra can be reviewed in the work by \citet{2015RAA....15.1182G}. Figure~\ref{fig:dwarf_vs_giant} illustrates the differences in spectral features between a typical dwarf and a giant with the same spectral type of M6. Prominent features for dwarfs include the \ion{K}{i} doublets near 7688\AA\  and \ion{Na}{i} near 8189\AA \citep{2009ssc..book.....G}. The hydride band of CaH between 6800 and 7000\AA~ is stronger in dwarfs than in giants; the bands for CrH and FeH around 8700~\AA ~ are visible in dwarfs. In comparison, giants have a more suppressed TiO absorption band to the red side of the \ion{Na}{i} doublets and stronger \ion{Ca}{ii} triplets at 8498~\AA, 8542~\AA, and 8662\AA. Moreover, giants exhibit a gentler slope between 8500~\AA ~ and 9000~\AA ~ compared to dwarfs that otherwise have molecular absorption due to TiO, VO, CrH, and FeH.

To select the sample of ultracool dwarfs, we applied visual inspection for all the 11,412 late M type spectra based on the above features, including both dwarfs (3764 dMs) and giants (7648 gMs) from the LDR7 official catalogue.  After the inspection of the spectra,  we found there are nine ``real ultracool dwarfs'' that had been misclassified as giants by the LAMOST pipeline and released in the giant catalogue; thus, we drew them back to our ultracool dwarf sample. For the LDR7 released dwarf sample, we excluded spectra with S/N lower than 2 in the z band, accounting for about four percent, as well as those of earlier M-type spectra that had been  misclassified as late-type by the LDR7 pipeline. We also removed the spectra that have features between dwarfs and giants, namely, those with obviously weaker \ion{K}{i} and \ion{Na}{i} lines than classical dwarfs, such as YSOs, or those having features other than classical dwarfs (caused by data reduction, or being the mixture of two or multiple very close stars). Finally, we got an ultracool dwarf catalog of 734 sources, which is described in the following subsection. We should note that this sample is conservative to ensure every source selected is indeed a real dwarf.

\subsection{Sample of ultracool dwarfs}
\label{sec:field}

\begin{figure}
\centering
\includegraphics[width=89.3mm]{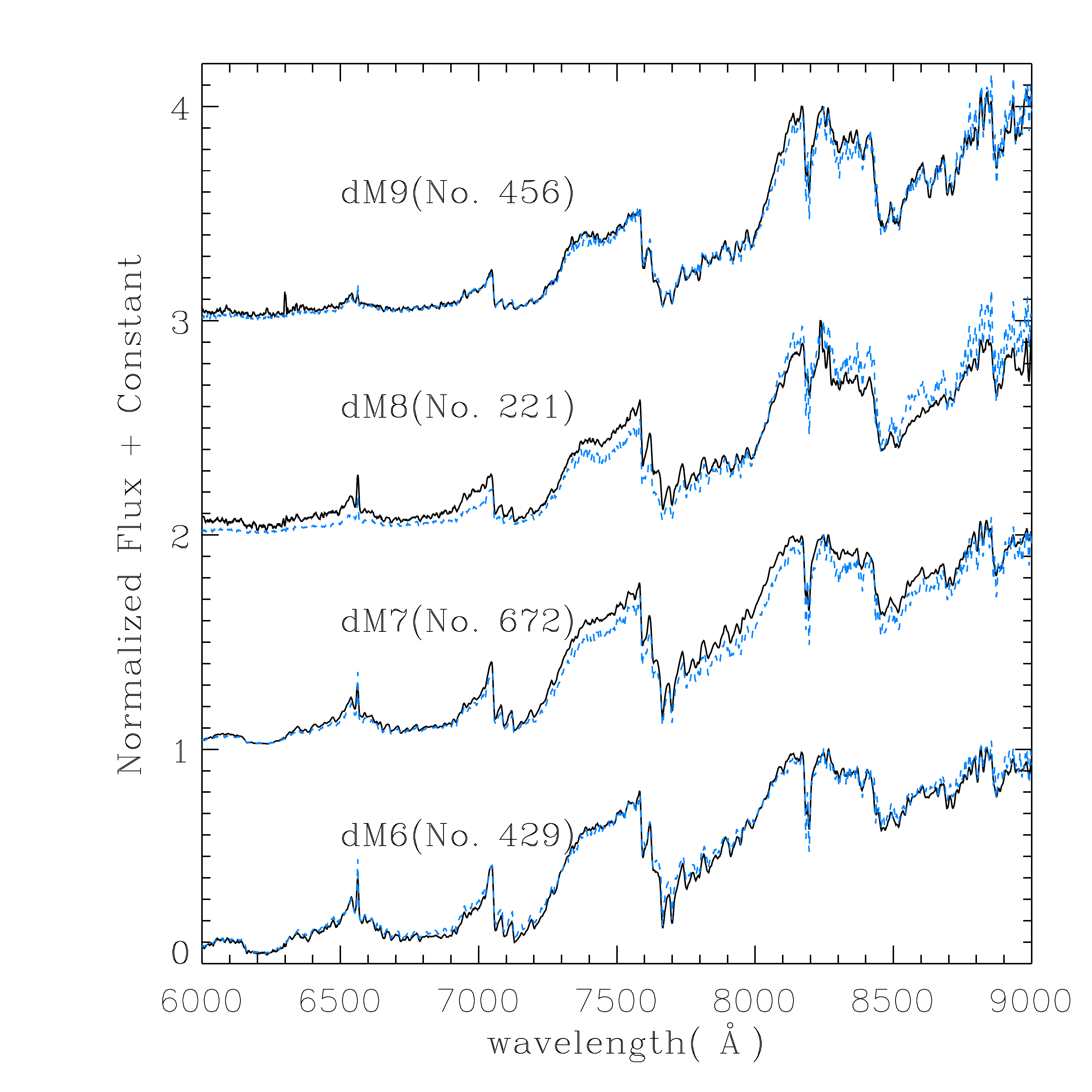}
\caption{\label{fig:mdwarfs4} Typical late-M dwarf LAMOST spectra (solid lines) in comparison to the Hammer template (blue dashed lines). Numbers given beside each spectral type corresponds to ID in Table~\ref{tab:BDs}. The S/N in SDSS $z$ band for No.~429~ $\sim$ 70, No.~627~ $\sim$ 174, No.~221 $\sim$ 30, No.~456~ $\sim$ 33.}
\end{figure}

The visually identified 734 late-type M dwarfs have spectral type equal to or later than M6 V, with one L0 included. Figure~\ref{fig:mdwarfs4} exhibits typical spectra of this sample. This sample is by no means complete in terms of sky coverage or flux but should form a reliable collection of the ultracool dwarf population in the LAMOST database. The observed spectra in our sample are consistent with the Hammer M dwarf templates, which have been established using the Sloan Digital Sky Survey \citep[here after SDSS,][]{2009ApJS..182..543A} optical spectra \citep{2007AJ....134.2398C,2014AJ....147...33Y}.

Table~\ref{tab:BDs} lists the 734 ultracool dwarfs that we identified using LDR7 spectra, including 577 of the dM\,6 and three gM\,6, 128 of the dM~7  and two gM~7 type, 16 of the dM~8 and one gM~8 type, three of the dM~9 and three of the gM~9 type, and one of the L~0 type.  Based on a cross-match with SIMBAD, we note that 625 out of the 734 objects had no spectral type listed in the literature prior to this work by LAMOST spectral observation.

\begin{table*}\centering
\scriptsize
\caption{Ultracool dwarfs in the sample}
\label{tab:BDs}
\begin{tabular}{@{}l@{~~}c@{~~}c@{~~}c@{~~}c@{~~}c@{~~~}c@{~~}c@{~~}c@{~~}c@{~~}c@{~~}c@{~~}c@{~~}c@{~~}c@{~~}c@{~~}c@{}}
\hline
No.&obsid & R.A.& Decl.& SpT1 & source $\textunderscore$ id & $\varpi$& $\Delta \varpi$ &$\mu_\alpha \cos\delta$& $\Delta \mu_\alpha \cos\delta$& $\mu_\delta$ & $\Delta \mu_\delta$ &RUWE&$G$ & $G_{BP}$ & $G_{RP}$ &SpT2 \\
 & & \multicolumn{2}{c}{(deg)} & & &\multicolumn{2}{c}{(mas)}& \multicolumn{2}{c}{(mas/yr)} &
            \multicolumn{2}{c}{(mas/yr)}& & \multicolumn{3}{c}{(mag)} &  \\ \hline
1&369905116&0.29067&7.30335&dM7&2745889793400449280&25.43&0.15&135.18&0.19&-78.93&0.09&1.461&16.333&18.274&14.390& -\\
2&472904167&0.64809&17.48399&dM6&2772949702272966144& - & - & -  & -  & -  & - & - &16.756&18.395&14.938& - \\
3&474610041&0.77652&39.20759&dM6&2881030840586539264&15.91&0.06&68.35&0.04&-56.65&0.04&1.009&16.574&18.716&15.222& -\\
4&592310145&1.82317&36.46536&dM6&2877146781401666816&14.19&0.07&16.77&0.08&-38.87&0.05&1.112&16.692&18.702&15.374& -\\
5&593009044&2.23202&49.31638&dM6&393621524910343296&67.76&0.04&352.67&0.03&202.42&0.03&1.103&14.385&16.769&12.998& -\\
6&472906086&2.45851&17.15581&dM6&2772737844422061440&28.13&0.07&-6.68&0.10&-77.52&0.08&1.005&16.340&18.788&14.941& -\\
7&354310106&4.60190&36.47204&dM6&2876595273240771840&20.78&0.05&174.36&0.04&-206.20&0.05&1.003&16.127&18.178&14.791& -\\
8&284610069&5.12607&33.08272&dM7&2863419584886542080&81.57&0.04&1075.61&0.04&-863.67&0.04&1.052&13.935&16.400&12.538& -\\
9&372110218&5.42199&36.32373&dM6&378382770487254912&10.47&0.11&-73.46&0.10&-26.90&0.10&1.726&16.640&18.522&15.275& -\\
10&614109230&6.12484&36.83748&dM6&378467643336605056&30.63&0.05&-132.67&0.04&-14.74&0.04&1.046&15.703&17.855&14.356& -\\
\vdots&      \vdots &   \vdots &   \vdots &    \vdots&   \vdots&       \vdots&    \vdots&    \vdots &    \vdots&    \vdots&
\vdots&       \vdots&     \vdots&    \vdots&     \vdots&    \vdots \\ \hline
\end{tabular}
\tablefoot{Column 1:\ running number. Column 2: LAMOST obsid. Columns 3-4: Right ascension(RA) and declination(Decl.). Column 5: LAMOST spectral type. Column 6: GEDR3 source$\textunderscore$ id. Columns 7-12: parallax, proper motion in R.A. and Decl. directions, and the corresponding uncertainty. Columns 13-15: \textit{Gaia} G BP and RP magnitudes. Column 16: RUWE from GEDR3. Column 17:  SpT from the literature according to SIMBAD. The electronic version of this table in entirety is available at http://paperdata.china-vo.org/YFW/aa\textunderscore table1.csv }
\end{table*}

The only one L dwarf, listed in Table~\ref{tab:BDs} as No. 556, with the designation $SDSS~J120430.38+321259.4$, has a SDSS spectral type of L\,0e in the optical \citep{2008AJ....135..785W} and M\,9.5 in the infrared \citep{2014ApJ...794..143B}. It was observed in 2012 during the LAMOST pilot survey but was not included in the release until LAMOST DR6 because of its low brightness. This L0 dwarf indicates the practical limit of the latest spectral type in LDR7.

\begin{figure*}
\centering
\includegraphics[width=160mm]{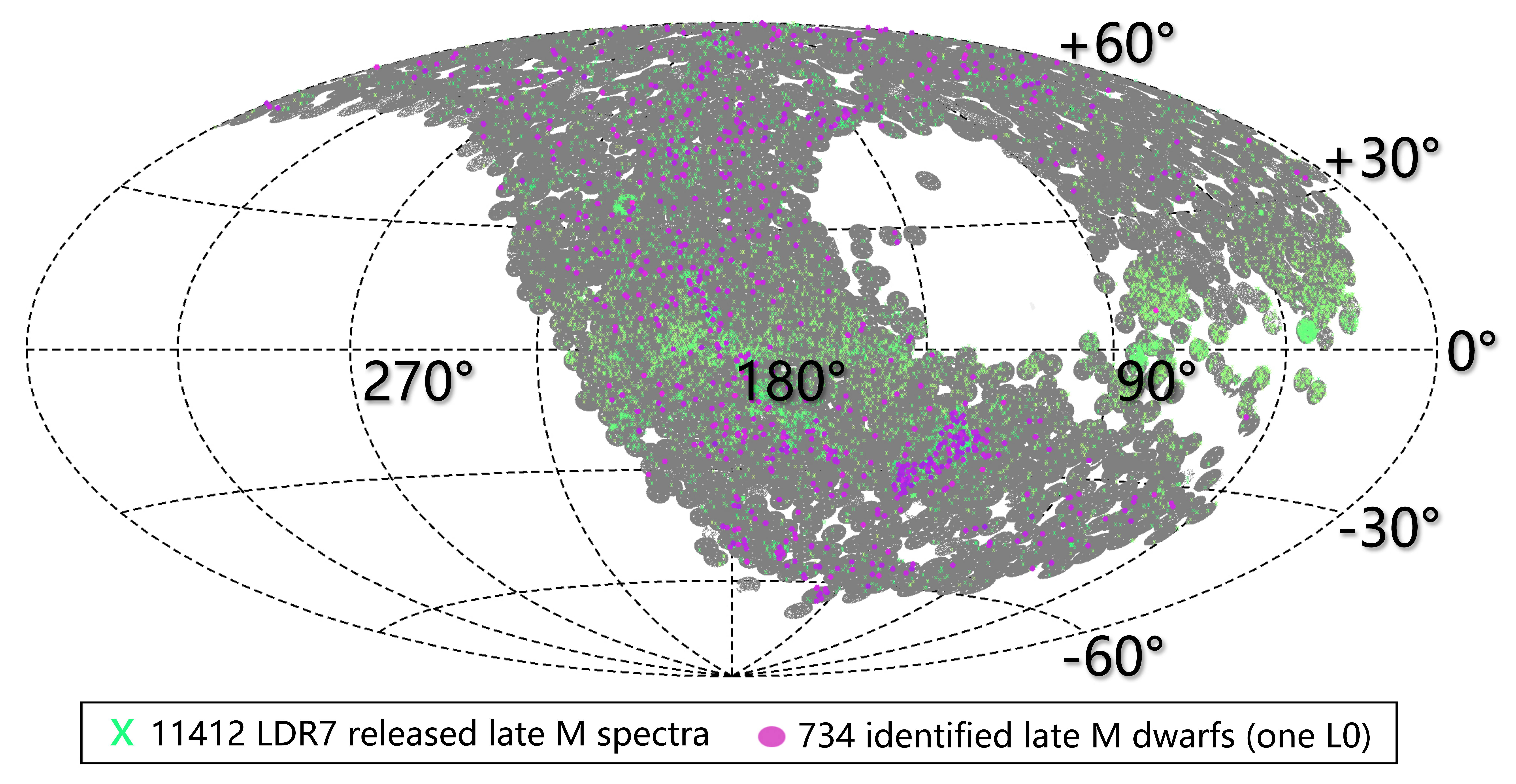}
\caption{\label{fig:rade} Sky distribution of  LAMOST DR7 released 11,412 late M stars  are shown in green crosses),  while the identified ultracool dwarfs are in light purple solid circles. The footprints for the LDR7 are plotted as grey big circles in the background.}
\end{figure*}

Figure \ref{fig:rade} shows the sky distribution of the ultracool dwarfs(light purple solid circles) and the 11,412 LDR7 released late M type spectra(in green crosses), the background grey circles are the footprints(consist of more than 4900 plates each plate covers 5 square degrees) of the LDR7, covering about 20000 square degrees of the northern sky. In order to test the completeness and ratio of the red faint objects observed, we downloaded the data from GEDR3 in the sky area of LDR7  and obtained about 130 million Gaia sources. We then empirically estimated the number of the late-type M dwarfs by the color cut of \textit{Gaia} $BP-RP>3$ and apparent magnitude cut of $G<19$, which result in 458 thousand objects. Using the absolute magnitudes of $G>9$, we got 143 thousand candidates, accounting for 0.11\% of the total. While for the approximately 10 million of the Gaia sources observed by LAMOST DR7, the pipeline classified 3,764 as late-type dMs, accounting for 0.03\%, which means the very low portion of late dMs due to telescope magnitude limits. Although most of the LDR7 released late-type M dwarfs are correctly classified as real M type,  many have low signal-to-noise ratio spectra or have fake spectral features. To avoid missing late dMs, we carefully check the spectral features for LAMOST released late dMs and gMs, and identified 734 relatively high S/N ultracool dwarfs later than M6, which ranging $5>BP-RP>2.63$ and absolute G magnitude between 8.94 and 16.71(with relative parallax error smaller than 20\%). Since our ultracool dwarf sample is composed of good quality spectra and populated primarily with solar metallicity single stars, we can obtain more reliable atmosphere parameters and use them to update LAMOST classification templates.

\section{GEDR3 photometry}
\label{sec:photometry}

The 11,412 late M spectra from LDR7 were cross-matched with the GEDR3 data within 5 arcsec as the previous section mentioned. In this section, we study the distribution of heliocentric distance and magnitude, $BP-RP$ versus $G-RP$ two-color diagram and CAMD diagram. Our ultracool dwarfs are very faint and red objects and represent the faintest stellar or substellar objects that LAMOST has ever observed. The heliocentric distance and \textit{Gaia} magnitude distribution show the practical limitation of the LAMOST.

To plot diagrams in this section, We restricted to objects with parallax that has relative uncertainty is smaller than 20\% \citep{1973PASP...85..573L,2018A&A...616A...9L}, and that all \textit{Gaia} $G, BP$, and $RP$ values are greater than zero. The parallaxes of these ultracool dwarfs are large and hence expected to be nearby, we, therefore, adopted the reciprocal of the GEDR3 parallax as the distance of each source and thus determined their absolute magnitudes.


\subsection{\textit{Gaia} magnitudes and distance distribution}
\label{sec:magnitudes}

\begin{figure}

\centering
\includegraphics[width=89.3mm]{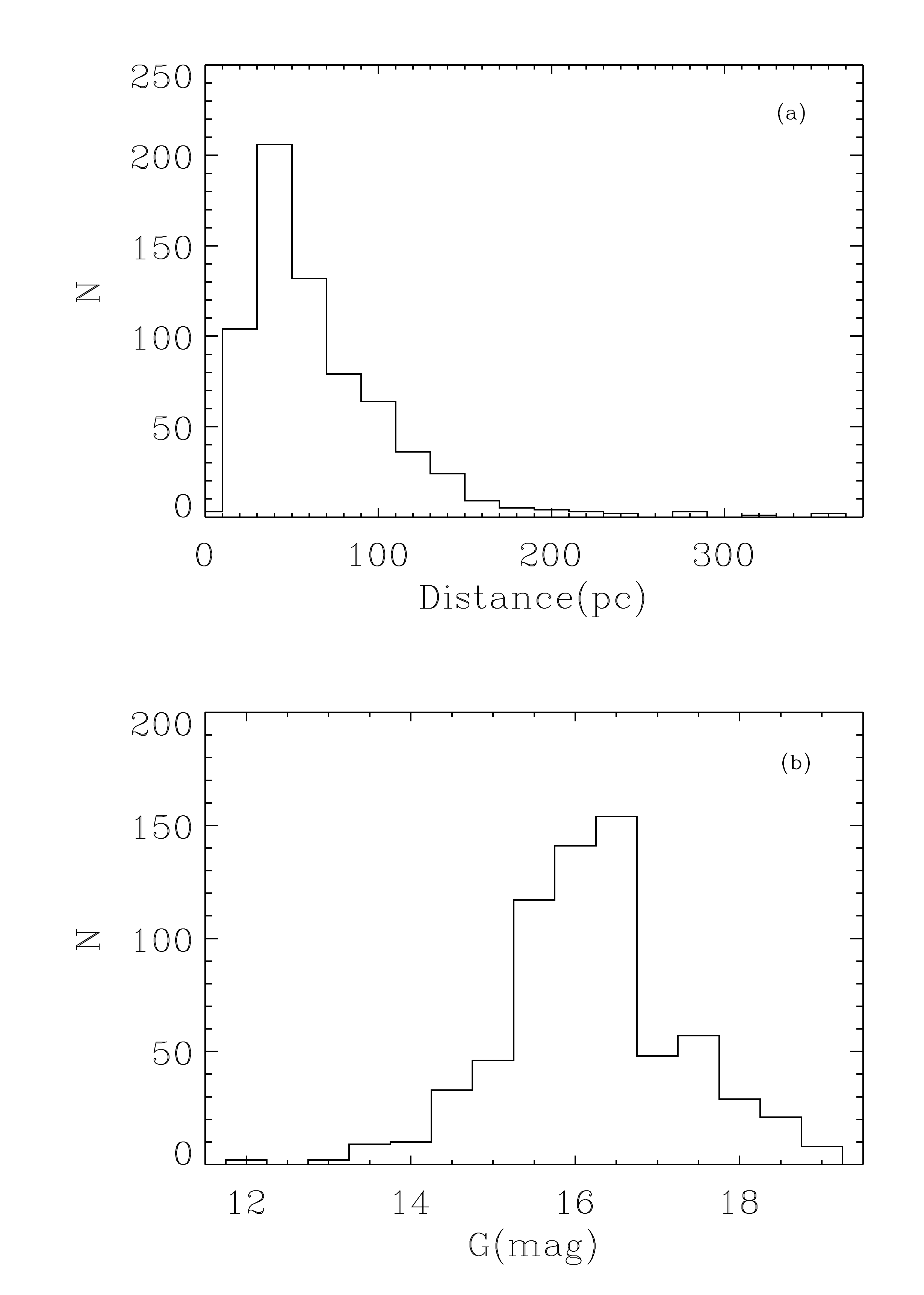}
\caption{\label{fig:dist_mag} Histograms of distance (panel a, reciprocal of the GEDR3 parallax, $binsize =20$~pc), \textit{Gaia} $G$ magnitude (panel b,$binsize=0.5$ mag).}
\end{figure}

There are 677 of the ultracool dwarfs that meet the above mentioned requirements. The distance distribution is shown in Figure~\ref{fig:dist_mag}(a), with the ultracool dwarf sample in black solid lines. Figure~\ref{fig:dist_mag}~(b) shows their \textit{Gaia} $G$ magnitude distribution.

Figure~\ref{fig:dist_mag}(a) indicates that the ultracool dwarfs are all within 360 pc. We notice that most of the ultracool dwarfs are within 100 pc ($\sim$ 83\%). The G magnitude range is from $\sim$ ~12 to $\sim$~19.134 mag, the peak is at around  $\sim$ 16 mag -- mean value of 16.206, corresponding to a spectral S/N measured in the equivalent SDSS $z$ band S/N of 2 to 432 with a median of 61. We note that our sample has spectra of good quality and is relatively convenient for model fitting or use as optical spectra templates.


\subsection{Testing the sample with colors and absolute magnitude}
\label{sec:hrd}

\begin{figure}
\centering
\includegraphics[width=89.3mm]{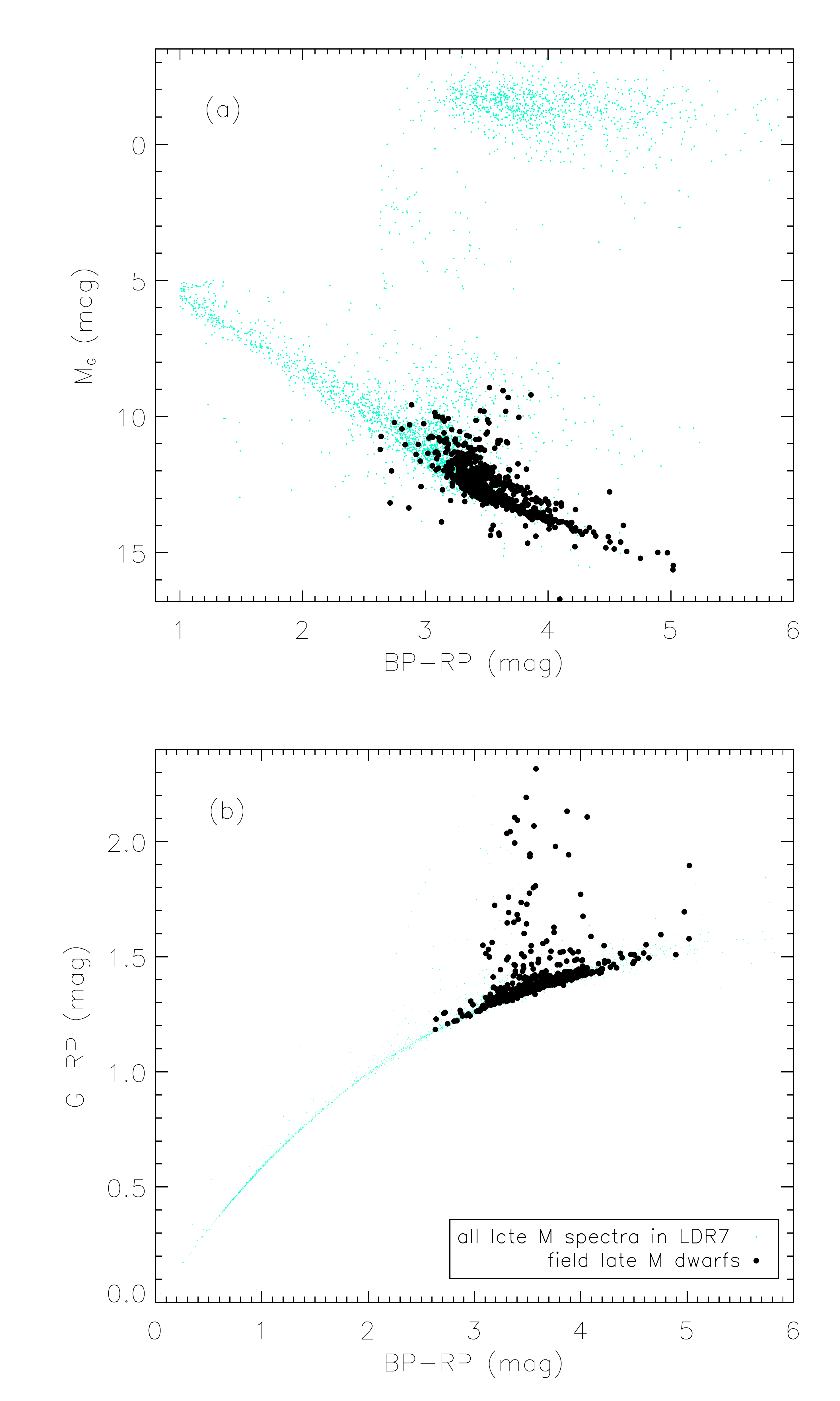}
\caption{\label{fig:tcdcmd}
            (a) $G$ absolute magnitude versus  $BP-RP$ diagram.  $G$ absolute magnitude computed from the inverse of the parallax with no correction of extinction. The entire LAMOST~DR7 late M sample is plotted in cyan dots, whereas the ultracool dwarfs are in black solid circles.
            (b)\textit{Gaia} $G-RP$ versus $BP-RP$ color-color diagram. Symbols are the same as in the top panel.}
\end{figure}

Figure~\ref{fig:tcdcmd} (a) exhibits the CAMD of $G$ absolute magnitude versus $BP-RP$, and (b) shows the two-color diagram of $G-RP$ versus $BP-RP$. For both (a) and (b) panels, the sources in Table~\ref{tab:BDs} are in black solid circles, and the 11,412 LDR7 released late-type M from LDR7 are in cyan dots.

In Figure~\ref{fig:tcdcmd}~ (a), the main sequence, giant and white dwarf loci are obvious for the 11,412  late M from the LDR7. Our ultracool dwarf samples are all located in the bottom right corner -- the loci of the reddest and coldest objects with $BP-RP$ between 2.63 and 5.00, G absolute magnitude between 8.94 and 16.71. It is reassuring to see that our ultracool dwarfs behave as expected, namely, they are consistent with being cool dwarfs. Although there seem to be white dwarfs below the upper main sequence among the 11412 released late M objects, the case is not always true. There are eight sources located at the loci of the white dwarf, but after checking their spectra, only three are DA white dwarfs and M dwarf binaries; whereas the other five do not show any obvious features of a white dwarf.

In panel (b), our ultracool dwarfs(black solid circles) locate at the top right -- the red and cold loci as well, consistent with the  \citet{2018A&A...619L...8R}  on $G-RP$ color. However, the $G-RP$ color has greater uncertainties than the $BP-RP$ colors, presumably due to the relatively shorter color baseline of the $G-RP$ \textit{Gaia} data. There are 49 sources with $G-RP$ greater than 1.5 when $BP-RP$ between 3 and 4.2, which is a fainter sub-sample with a mean G band magnitude of 16.825. They usually have RUWE (listed in table 1) value bigger than 1: the mean value is 3.28 and the median is 2.10. Greater RUWE values, that is, more than 1.4, indicates a significant possibility that there is an unresolved binary involved \citep{2020MNRAS.496.1922B}. The binary could be the cause of the peculiar $G-RP$ color. In Section 6.2, we show the $G-RP$ color for binaries in this paper. It is noteworthy that there are several high proper motion and wide separation binaries, which could lead to complexity and possible errors in the cross-matching process.

Our $BP-RP$ color appeared as expected, with normal distribution and without the problem \citet{2021A&A...649A...6G} remarked for the case of UCDs ( which have greater scatter on $BP-RP$ color caused by bigger uncertainty on $BP$ magnitude due to their red and faint nature) since our sample is restricted to brighter sources. There are 533 of the ultracool sample reported in the work of \citet{2021A&A...649A...6G}, where they reported the nearby stars within 100 pc selected by the \textit Gaia photometry and astrometry, and we report their spectral type in this work. Besides, our sample reported fourteen M6V to M8V  which are predicted by \citet{2018A&A...619L...8R} using \textit{Gaia  DR2} and 2MASS data. \citet{2016A&A...589A..49S} and \citet{2019A&A...623A.127A} reported 40,000 late-type M select by photometry, but only three objects in this sample overlap with our sample; this is probably because their samples are much fainter.

\section{Spectroscopic analysis}
\label{sec:spectrascopic}

 Mass or age is needed to ascertain their sub-stellar nature and neither quantity is readily available. We hence relied on other criteria such as the \teff,~ \logg,~\feh~and strength of the lithium line, all available in the catalog and spectra of LDR7, as proxies of mass and age.

\subsection{Atmospheric parameters}
\label{sec:parameter}

Stellar atmosphere parameters \teff~, \logg,~ and \feh~ for M type objects in LDR7 are provided by LAMOST Stellar Parameter Pipeline for M stars \citep[hereafter LASPM,][]{2021RAA....21..202D}. It minimizes $\chi^2$ between the observed low-resolution spectra and a linear combination of five best-matching grids in the wavelength range of 6000 to 8800 \AA~ with some specific contamination of the earth atmosphere masked \citep{2021RAA....21..202D} to derive the stellar parameters. The LASPM does not provide stellar parameters for spectra with S/N in the SDSS $i$ band bellow 5. The grids are from  BT-Settle CIFIST2011 grids \citep{2011ASPC..448...91A,2012EAS....57....3A}. The CIFIST2011 model is calculated from the radiative transfer code named ``PHOENIX'' with the solar abundance from \citet{2011SoPh..268..255C}, considering dust formation, gravitational settling, and convection. The line lists used are introduced briefly in \citet{2021RAA....21..202D}. The coverage of \teff,~\logg~ and \feh are 300 to 8000\rm{K}, 0.0 to 0.6 \rm{dex,} and -2.5 to 0.5 \rm{dex} with steps of 100 \rm{K}, 0.5 \rm{dex,} and \rm{0.5,} respectively.

The measured stellar parameters of the ultracool dwarfs are listed in Table \ref{tab:parameters}. Letters ``H, M, and L'' in column 10 refer the first letter of "high," "medium,'' and ``low,'' indicating the strength of the lithium line, with ``H'' the strongest and ``L'' the weakest, based on visual checking of the lithium line. The label ``N'' indicates the absence of the lithium line. We did calculate the equivalent width, but it is a weak feature and the flux near 6708 \AA~  is often rather noisy, so we do not report the value here.

\begin{table*}
\scriptsize
\caption{Radial velocity and atmosphere parameters in LDR7.}
\label{tab:parameters}
\centering
\begin{tabular}{ccc ccc ccc ccc}
\hline
R.A.& Decl.& RV& \teff & $eT_{eff}$ & \logg & e\logg & \feh & e\feh& Li& S/N $i$ & S/N $z$ \\
   \multicolumn{2}{c}{(deg)} & (\kms) &\multicolumn{2}{c}{(K)} & \multicolumn{2}{c}{(dex)}&\multicolumn{2}{c}{(dex)} &  & & \\ \hline
0.29067&0.29067&11.85&2960&87&4.80&0.14&-0.20&0.10&N&80&106\\
0.64809&0.64809&-20.56&3045&87&4.21&0.15&-0.27&0.10&N&69&95\\
\vdots&      \vdots &   \vdots &   \vdots &    \vdots&    \vdots&    \vdots&    \vdots&    \vdots&\vdots& \vdots&\vdots \\
15.77892&15.77892&22.22&3053&108&4.61&0.17&-0.23&0.12&N&13&19\\
15.88307&15.88307&-25.09&3063&97&4.88&0.16&-0.18&0.11&N&78&97\\
15.93150&15.93150&3.06&3091&75&4.30&0.13&-0.23&0.09&N&84&110\\
15.93207&15.93207&3.51&3081&100&4.72&0.17&-0.09&0.12&N&63&79\\
16.16718&16.16718&-0.17&3004&105&4.87&0.17&-0.25&0.12&L&57&82\\
 \vdots&      \vdots &   \vdots &   \vdots &    \vdots&    \vdots&    \vdots&    \vdots&    \vdots&\vdots &\vdots&\vdots \\ \hline
\end{tabular}
\tablefoot{The parameters in LDR7. Columns 1-2: Position from LDR7 in epoch J2000. Column 3: RV. Column: 4-9:\ effective temperature, surface gravity, and metallicity, each parameter with their uncertainty followed. Column 10 label indicates if there is a lithium absorption line at 6708 $\AA$ by ``H/M/L' '(with the strength of the lithium at ``H'' as the highest level, ``M'' the medium, and "L" the weakest) or ``N''(indicating no obvious lithium line). Columns 11 and 12:\ S/N in the SDSS $i$ and  $z$ band. Here we show only a portion of this table, and it is published in its entirety in the online machine-readable format at
http://paperdata.china-vo.org/YFW/aa \textunderscore table2.csv}
\end{table*}

We present the atmospheric parameters in Table \ref{tab:parameters}. The effective temperature spread is from 2600~K to 3300~K, while the mean temperatures for each SpT are: M6~$\sim$~ 3018~K, M7~$\sim$~ 2917~K, M8~$\sim$~ 2775~K, and M9~$\sim$~ 2675~K. The mean temperatures cool down as the spectral type gets latter from M6 V to M9 V, and are roughly consistent with the literature that adopts atmospheric grid fitting with low or mid to high-resolution spectra  \citep{2013A&A...556A..15R,2019ApJ...879..105L} within 200K, but they are usually higher than the literature. We note there are only a few M8 and M9 dwarfs in both literature and our sample which makes the comparison difficult. If we make a comparison with the results of \citet{2005nlds.book.....R} or \citet{2009ApJ...698..519K}, our mean temperatures are higher by about 350~K for each same SpT for late M dwarfs. One reason for this bias might be caused by the saturation of TIO absorption band in spectra later than M6. The TiO band is sensitive to temperature and weak for high temperatures, and then starts to saturate to the cool end  \citep{2016A&A...587A..19P}. This TiO behavior was not properly described in earlier versions of the PHOENIX models. The surface gravity varies from 3.5 to 5.5 dex with a mean of 4.43 dex. The measurement of surface gravity from the LASPM is broadly consistent with our expectations. And the metallicity varies from -0.99 to +0.38. The mean metallicity is $<\feh>~ = -0.33$, consistent with Galactic disk dwarfs according to \citet[$<\feh>~=-0.4$,][]{1995AJ....110.2771W}.

Thus, we give several cautions for our derived stellar parameters.\\

Firstly, the LASPM assumed the giants(with a spectral type of ``gM'') have a surface gravity between 0 and 4.00 dex, while the dwarfs (with a spectral type of ``dM'' ) have between 3.5 and 5.5. If a dwarf is misclassified as a ``gM'' it is likely this dwarf will have \logg~ value that is smaller than 3.5. In our ultracool dwarf sample, there are 16 objects that have a surface gravity smaller than 3.5, but we checked their spectra and we know that they have prominent \ion{K}{i} doublets near 7688\AA, and \ion{Na}{i} near 8189\AA. So, for these objects, we set their surface gravity at -9999.99.

 Next, we know that LASPM assumed the giants (with spectral type of ``gM'') have surface gravity between 0 and 4.00 dex, while the dwarfs (with a spectral type of ``dM'' ) between 3.5 and 5.5. If a dwarf is misclassified as ``gM,'' it is likely this dwarf will have \logg~ smaller than 3.5. In our ultracool dwarf sample, there are 16 objects that have surface gravity smaller than 3.5, but we checked their spectra and we know that they have prominent \ion{K}{i} doublets near 7688\AA, and \ion{Na}{i} near 8189\AA. So, for these objects, we set their surface gravity at -9999.99.

Lastly, although the values for the stellar parameters are statically consistent with the expectations, it is better to check the spectra before adopting the parameters because LASPM, which is a pipeline aimed at compute parameters for a large number of objects, does not guarantee the accuracy of parameters for each individual object. In Table \ref{tab:parameters}, the last two columns list the S/N for consideration.

\subsection{Lithium diagnosis}
\label{sec:lithium}

Given the uncertainty in \teff,~\logg~ and \feh, we did not attempt to estimate the mass of each of our ultracool dwarfs by connecting the atmospheric model to evolutionary ones. Instead, we can use a more direct proxy by looking for the presence of the lithium absorption line at 6708~\AA. Given the low dispersion of the LAMOST DR7 spectra, we did not measure the equivalent width. Generally, the lithium test for BDs, including analysis of the equivalent width, requires an 8 to 10-meter telescope and mid-to-high dispersion spectroscopy \citep{1995Natur.377..129R,1999AJ....118.2466M,2008ApJ...689.1295K}.

\begin{figure}
\centering 
\includegraphics[width=89.3mm]{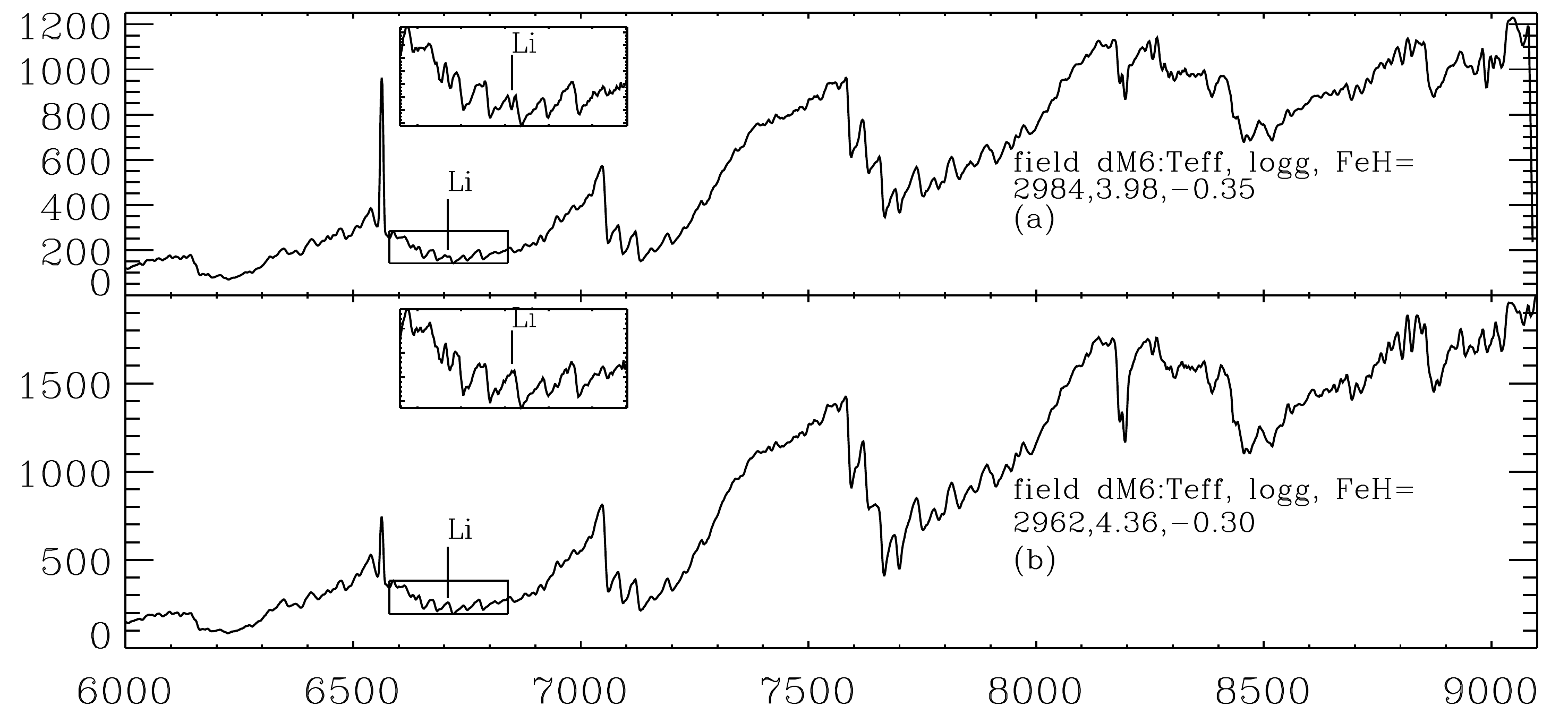}
\caption{ \label{fig:lithium} Comparison of two spectra based on their 6708 \AA ~lithium lines: One with the lithium line and another without. Panel (a) shows an M6 spectrum with a lithium absorption line. An M6 spectrum without lithium is shown in panel (b) for comparison. The small windows in each panel zoom in on the region that containing the 6708~\AA lithium line. The atmosphere parameters from the LAMOST 1D pipeline are labeled.}
\end{figure}

The lithium line for our ultracool dwarfs was checked. Figure \ref{fig:lithium} compares an ultracool dwarf showing the lithium line and another one without the lithium line, each with an inset zooming into the wavelength range about the lithium line at 6708~\AA. In Table~\ref{tab:parameters}, as mentioned in the previous subsection, the 10th column labels ``H/M/L'' for the targets showing the lithium line, indicating the strength of high, medium, or weak, whereas ``N'' marks those with no detection of the line. There are 77 out of the 734 ultracool dwarfs with a prominent lithium absorption line at 6708 \AA. We further explored the lithium line using the LAMOST single exposure spectra to exclude the impact of one bad single spectrum and to make sure that none of the above-mentioned lithium lines is caused by noise.  

There is roughly 10\% of our ultracool dwarf sample exhibiting the lithium feature, which is consistent with the study from  \citet{2009ApJ...705.1416R}. The presence of a lithium line in a cool M dwarf with an age older than a few hundred Myr is firm evidence of mass below 65~$M_J$, namely, below the hydrogen-burning limit; therefore, it is a brown dwarf \citep{2009ApJ...705.1416R}. However, lithium depletion is expected to take about 100 Myr to reach completion. Definitive inference from the presence of lithium requires that reliable ages are established for all objects, which is not available especially for the field objects. Thus, it is safe to say that these 77 ultracool dwarfs are very probably brown dwarfs and that these lithium detections represent the first automated survey mode discoveries of lithium, establishing a significant new sample of lithium detections.


\section{Kinematic diagnosis
\label{sec:kinematic}}

Our ultracool dwarfs are all within 360~pc, namely, they are in the solar neighborhood and likely to be Galactic disk members. However, there is a possibility that they may have a halo origin and will have a relatively high space velocity.

In this section, we study their kinematics. We calculate the $UVW$ space motion and Galactic position using the RV (listed in Table \ref{tab:parameters}) from LDR7 1D pipeline and the astrometric data from GEDR3. The position, namely, the Galacticocentric distance and Galactic height, suggests that they are disk objects. With their LSR-corrected $UVW$ velocity, we plotted our ultracool dwarfs on the Toomre diagram. In addition, we calculate the probability ratios of the thin-disk-to-thick-disk as well as that of the thick-disk-to-halo. In short, we conclude most of them are thin-disk objects. Using the thin-disk objects, we calculate the mean and scatter of the $UVW$ velocities to infer their kinematic ages.

\subsection{Space position and velocity
\label{sec:compute}}

To ensure that the data we used are reliable, we excluded the data with relative uncertainty on parallax greater than 20\%. Considering the RV, because of the typically broad spectral features and relatively poor S/N for M type spectra (especially for late M types) the RVs from low-resolution spectra provided by LDR7 for our ultracool dwarfs are potentially not reliable and we should exclude the unreliable RVs.

\begin{figure}
\centering
\includegraphics[width=89.3mm]{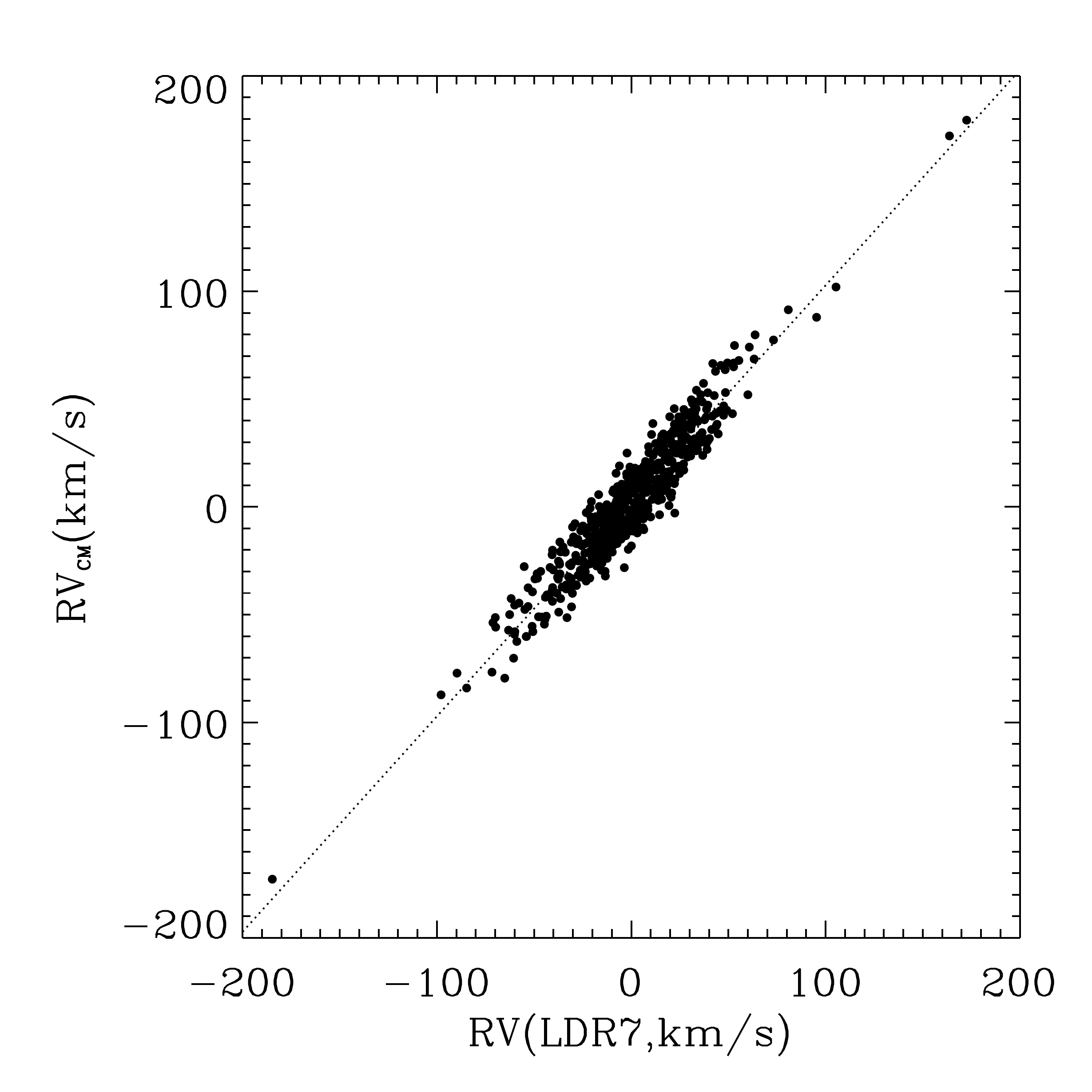}
\caption{\label{fig:rv_compar} Comparison of the LDR7 RV versus $RV_{CM}$ after a 3 sigma clip on the residuals of the two sets of RVs. The x-axis is the LDR7 RV and y-axis $RV_{CM}$. The ultracool dwarfs in our sample are shown in black solid circles.}
\end{figure}

Thus, we measured the RV based on cross-correlation with M type spectral templates from \citet{2015RAA....15.1154Z} for comparison with the RVs that come from LDR7. Our RV measurements are called $RV_{CM}$ hereafter. Then we applied an iterative three-sigma clip to residuals of the two sets of RV to cut out unreliable RVs. Figure \ref{fig:rv_compar} shows the final RV comparison after iteration with a three-sigma clip. The x-axis is from LDR7 and the y-axis is from our $RV_{CM}$(in black solid circles) the standard deviation is 9.96 \kms. We adopted the LDR7 RV and 10\% of RV for uncertainty, which is a relatively conservative value for uncertainty.
The final adopted RV from LDR7 is consistent within three sigmas of $RV_{CM}$ but the two RVs have an offset of about 3.0 \kms. After the exclusion of all unreliable measurements, there are 541 sources left in our sample.

Using the five GEDR3 astrometric parameters (R.A., Decl., proper motion, and parallax) and the LDR7 RV, the $U V W$ space motion, then Galactocentric distance ($R$) and Galactic height ($Z$) are computed based on the following assumptions: 1) space motion with the U positive towards the Galactic center, V positive toward the Galactic rotation, and W positive to the Galactic north pole; 2) the distance between the earth and Galactic center is $R_{\bigodot}=8.2$ kpc \citep{2016ARA&A..54..529B}; 3) the motion of the local standard of rest (hereafter LSR) is 238 \kms, and velocity of the sun with respect to the LSR is $[U_{\bigodot}, V_{\bigodot}, W_{\bigodot}]=[14.0,12.24,7.25]$ \kms \citep{2010MNRAS.403.1829S,2012MNRAS.427..274S,2016ARA&A..54..529B}; and 4) the distribution of equatorial coordinates, proper motion, and parallax is a multivariate Gaussian\citep{2019MNRAS.490..157M,2021ApJS..252....3L}. The derived $U V W $, Galactocentric distance($R$) and Galactic height($Z$) are listed in Table \ref{tab:kinematic}.

\begin{table*}
\scriptsize
\caption{Kinematic parameters. }
 \label{tab:kinematic}
\centering
\begin{tabular}{ccc ccc ccc}
\hline
R.A.& Decl.& $U$ & {V} & {W} &{Z}&{R}& {TD/D}& {TD/H}  \\
   \multicolumn{2}{c}{(deg)}&\multicolumn{3}{c}{(\kms)} & \multicolumn{2}{c}{(pc)}& \multicolumn{2}{c}{} \\ \hline
0.29067&0.29067&$-16.47^{+0.17}_{-0.17}$&$-14.41^{+0.69}_{-0.70}$&$-22.55^{+0.87}_{-0.98}$&$-6.58^{+0.19}_{-0.19}$&$8204.83^{+0.08}_{-0.09}$&0.01&5157\\
1.82317&1.82317&$12.80^{+1.34}_{-1.33}$&$-37.79^{+3.13}_{-3.08}$&$3.72^{+1.59}_{-1.64}$&$-5.34^{+0.15}_{-0.16}$&$8225.22^{+0.13}_{-0.13}$&0.01&4663\\
2.23202&2.23202&$-21.07^{+0.36}_{-0.41}$&$-17.74^{+0.83}_{-0.76}$&$11.62^{+0.19}_{-0.22}$&$21.71^{+0.01}_{-0.01}$&$8206.28^{+0.00}_{-0.00}$&0.01&4965\\
2.45851&2.45851&$8.02^{+0.14}_{-0.14}$&$-10.95^{+0.41}_{-0.40}$&$-4.75^{+0.43}_{-0.43}$&$0.07^{+0.06}_{-0.06}$&$8208.28^{+0.04}_{-0.04}$&0.01&5259\\
5.12607&5.12607&$-32.29^{+0.54}_{-0.52}$&$-60.11^{+1.06}_{-1.12}$&$-44.20^{+0.71}_{-0.67}$&$19.01^{+0.00}_{-0.00}$&$8204.61^{+0.01}_{-0.01}$&0.28&2186\\
  \vdots &   \vdots &   \vdots &    \vdots&    \vdots&    \vdots&    \vdots&    \vdots&\vdots \\ \hline
\end{tabular}
\tablefoot{The kinematic parameters. Columns 1-2 show the position from LDR7 in J2000, columns 3-5 are space motion U V W with a unit of \kms. Columns 6-7 show the Galactic z and distance R with units in pc. The last two columns list the probability of Galactic membership, TD/D is the probability ratio of the thick disk to the thin disk, while TD/H indicates the probability ratio of the thick disk to the halo. Here is only a portion of this table, this table is published in its entirety in the online machine-readable format at http://paperdata.china-vo.org/YFW/aa \textunderscore table3.csv }
\end{table*}

\subsection{Galactic population analysis
\label{sec:disk}}

\begin{figure*}
\centering
\includegraphics[width=160mm]{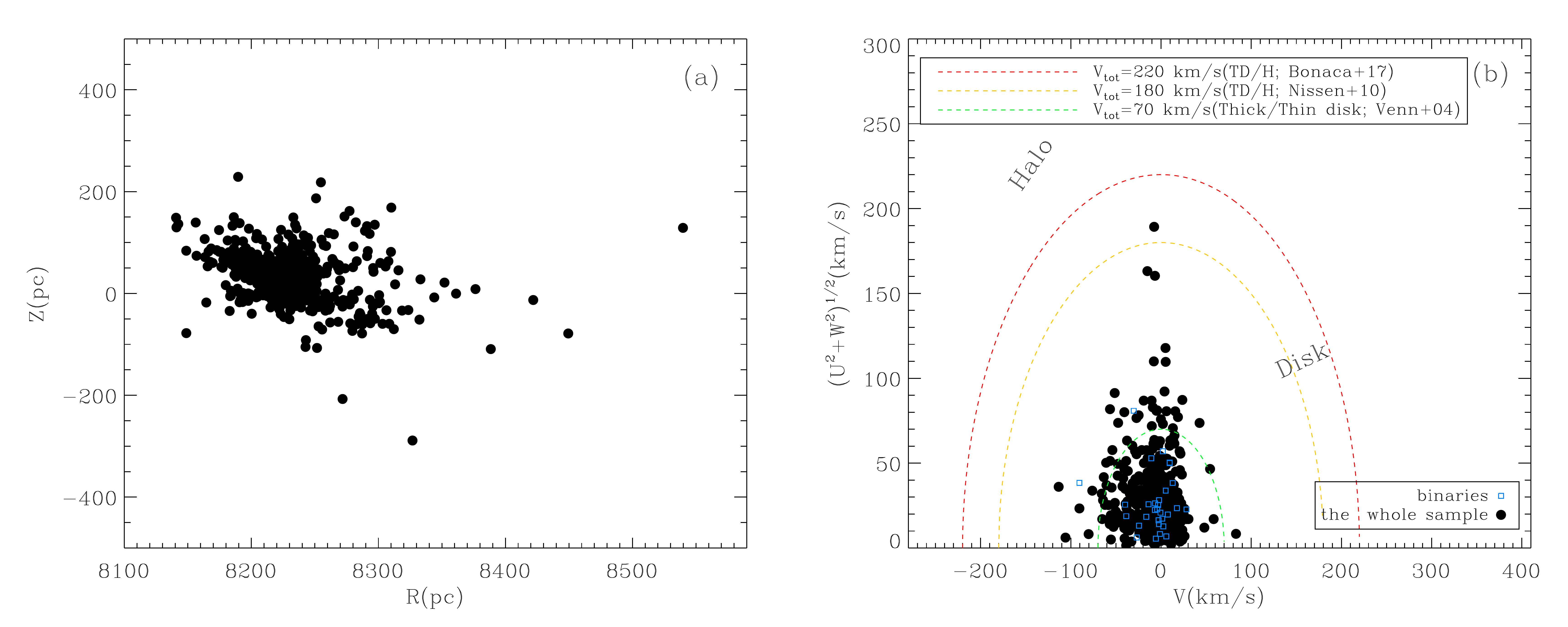}
\caption{\label{fig:toomre} Galactic R versus Z diagram (a). Toomre diagram for our objects (b). For both plots (a) and (b), our ultracool dwarfs are in black solid circles. The blue squares are objects in binary systems, The green dashed curve shows the thick-thin disk limit \citep{2004AJ....128.1177V} of total velocity. The blue and yellow dashed curve indicates the thick and halo limit from \citet{2010A&A...511L..10N} or \citet{2017ApJ...845..101B}.}
\end{figure*}

We show in Figure~\ref{fig:toomre} the derived $UVW$ velocities relative to the LSR and the Galactic position $R$ and $Z$. The black solid circles indicate the objects in our sample, in panel (b), the blue squares show objects in binary systems. The Galactocentric distance and Galactic height are shown in plot (a). From this figure, their Galactic heights are between $\sim$ ~ -289~ and +229~pc and the Galactic distances are between 8.1 and 8.5~kpc, which means they are the nearby local sample dominated by thin-disk objects. Further, we judge their Galactic membership using the Toomre diagram in plot (b) of Figure
\ref{fig:toomre}. The dashed lines in different colors are the total velocity limits between disk and halo from the literature. The green line indicates the thick-thin disk limit\citep{2004AJ....128.1177V}, the yellow and red are the thick disk-halo limit from \citet{2010A&A...511L..10N} and \citet{2017ApJ...845..101B}, respectively. Adopting either \citet{2010A&A...511L..10N} or \citet{2017ApJ...845..101B}, we see that the ultracool dwarfs are disk members in our sample and most of our sources are located within the thick-thin disk limit. There are 46 out of the 541 single objects located beyond the thin-thick disk limit but within the thick disk-halo limit; they are thick-disk objects, leaving 495 as thin-disk objects. The blue squares in Figure \ref{fig:toomre} panel (b) remind us of the binaries that we need to consider when doing further analysis in Section \ref{sec:distr_age}.

We studied the Galactic membership through thick-disk-to-thin-disk (hereafter TD/D) and thick-disk-to-halo (hereafter TD/H) probability ratios, which are calculated considering the distribution of thin-thick-halo components in velocity as well as the observed fractions of each population in the solar neighborhood following \citet{2003A&A...410..527B}. The probability for each object is listed in Table \ref{tab:kinematic}. Two of our objects have the TD/H probability of 0.06 and 0.09 (less than 0.1), but according to their Galactic position (Figure~\ref{fig:toomre}), we conclude they are disk objects. Adopting a TD/D probability smaller than 0.1, 491 of the 541 objects are within the thin disk. In principle, the velocities of the objects in binary (as shown in Fig. \ref{fig:toomre}) or multiple systems will be affected by their companions, so we did not include them in the following kinematic study. In Section \ref{sec:cpm}, we further discuss these binaries.

We conclude that most of our ultracool dwarfs are thin-disk objects, while a small portion of them are thick-disk objects that are consistent with the metallicity analysis (see Section \ref{sec:parameter}). In the next sub-section, we compute the $UVW$ average value with respect to the Sun and the dispersion for our ultracool dwarfs, so as to estimate the kinematic age. For further study, we require the thin-disk objects to be judged as thin-disk stars by both the Toomre diagram and the probability ratio according to \citet{2003A&A...410..527B} method. There are 477 single thin-disk objects that meet our requirements. Correspondingly, there are 70 single thick-disk objects.

\subsection{Velocity distribution and kinematic ages
\label{sec:distr_age}}

\begin{figure*}
\centering
\includegraphics[width=160mm]{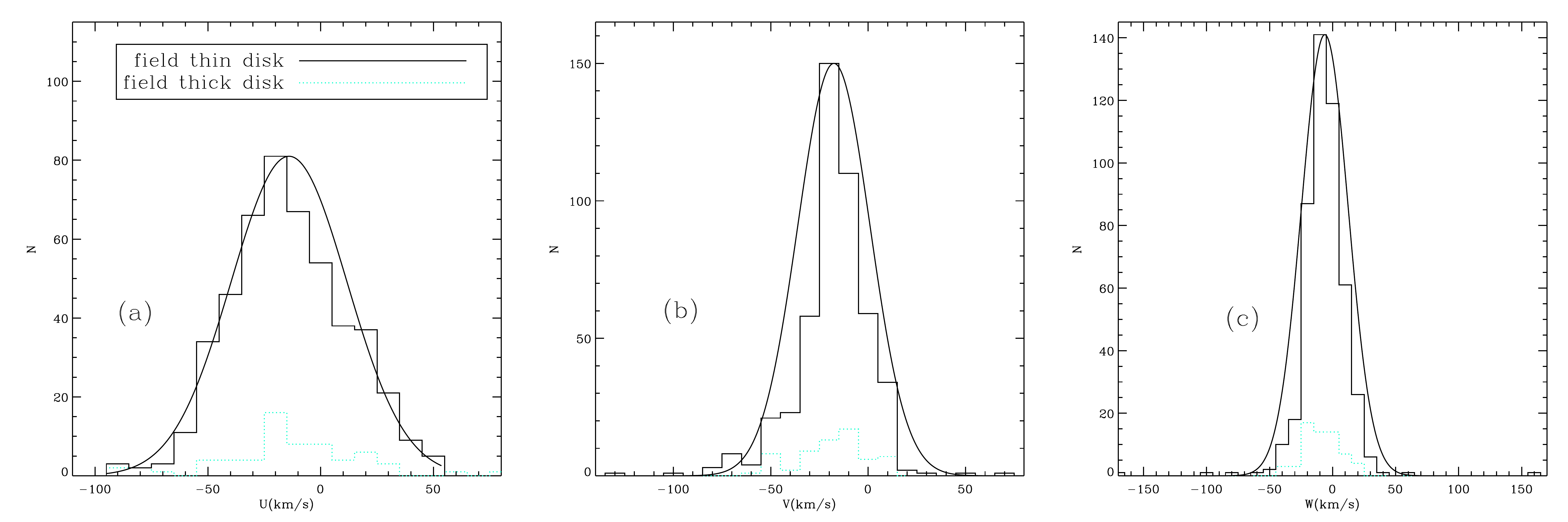}
\caption{\label{fig:uvw1} Histograms of $UVW$ velocities for the ultracool dwarfs. Objects in binary systems are deleted. Black solid histograms over-plotted with the fitted Gaussian curve for the 477 thin-disk objects. Cyan dotted bars show the 70 thick-disk objects. Bin size is 10 \kms~for all three panels.}
\end{figure*}

Figure~\ref{fig:uvw1} shows the distribution of the $UVW$ velocities for the thin-disk objects over-plotted with a Gaussian fit (in black  ) and the thick-disk ones (in dotted cyan bins). The binaries are not included in this figure to ensure that the objects shown are all singles to the best of our knowledge. We find that the velocities of the thin-disk objects in our sample have a prominent Gaussian profile and the velocity dispersion is very small. While for the thick disk components, the $W$ velocity shows a prominent Gaussian peak in Figure ~\ref{fig:uvw1} panel (c), but the distribution of $U$ and $W$ velocity does not look like a Gaussian distribution. This may be caused by the relatively small number of only 70 plotted. Hence, we did not carry out further kinematic analyses for the thick-disk objects in the following sections.

The mean and sigma of the $UVW$ for singles(477 thin-disk objects) are presented below in Table \ref{tab:kine_distr}. Our velocity data in this sub-section are heliocentric, so the average value of the sample reflects the anti-motion of the solar system relative to these dwarfs. Using our objects for reference, we find $(U,V,W)_{\bigodot}=[14.02^{+1.20}_{-1.21},17.60^{+0.71}_{-0.71},6.19^{+1.21}_{-1.20}]$~\kms; these average velocities agree within 3$\sigma$ of previous literature results for the solar motion; for instance, \citet{2018PASP..130f4402W} studied the $UVW$ velocity of about 200 L and T dwarfs, and found the solar motion of $(U,V, W)_{\bigodot}=[7.9\pm 1.7,13.2\pm 1.2,7.2\pm1.0]$~\kms~. The $U$ and $W$ velocity of LSR we derived are consistent within 3 $\sigma$ with the adapted values in section \ref{sec:compute}: $(U,V, W)_{\bigodot}=[14.0,12.24,7.25]$\kms \citep{2010MNRAS.403.1829S}, while the $V$ velocity is slightly bigger. The $V_{\bigodot}$ is subject to controversies with the value varying from 1 to 21 \kms \citep{2012ApJ...759..131B,2017A&A...605A...1R,2021A&A...649A...6G}.

\begin{table*}
\scriptsize
\caption{\label{tab:kine_distr}Velocity distributions of thin-disk objects and their kinematic ages.}
\centering
\begin{tabular}{c ccc ccc ccc} \\ \hline
     & $<v>$& low err & high err &$\sigma_v$ & low err  & high err& Age & low err & high err \\
   &\multicolumn{3}{c}{(\kms)} &\multicolumn{3}{c}{(\kms)} & \multicolumn{3}{c}{ (Gyr)} \\ \hline
$U$ &  -14.03 &1.20&1.20&\multicolumn{6}{c}{ } \\
$V$& -17.60 & 0.71 &0.71&\multicolumn{6}{c}{ }  \\
$W$& -6.19 &  1.21&1.20& \multicolumn{6}{c}{ }  \\
$\sigma_U$& \multicolumn{3}{c}{ } &25.71 &1.17 & 1.18 & \multicolumn{3}{c}{ }  \\
$\sigma_V$&\multicolumn{3}{c}{ } &18.53& 0.89 &0.89 & \multicolumn{3}{c}{ }  \\
$\sigma_W$&\multicolumn{3}{c}{ } &18.60 & 1.43 & 1.37& \multicolumn{3}{c}{ }  \\
$\sigma_{tot}$&\multicolumn{3}{c}{ } &36.75 & 4.21 &4.07& \multicolumn{3}{c}{ }  \\
Age&\multicolumn{6}{c}{ } &2.97&0.90&1.06 \\  \hline
\end{tabular}
\tablefoot{Distribution of the $UVW$ velocity and their kinematic age.}
\end{table*}

The standard deviation:$[\sigma_{U},\sigma_{V},\sigma_{W}]=[25.71^{+1.17}_{-1.18},18.53^{+0.89}_{-0.89},18.60^{+1.43}_{-1.37}]$ and total standard deviation: $\sigma_{v}=36.75^{+4.21}_{-4.07}$ are all listed in Table \ref{tab:kine_distr}. Our velocity dispersion is consistent with literature results that focus on BD samples, for instance, (23.0, 15.8, 12.2) from \citet{2018PASP..130f4402W}, (25, 23, 20) \kms from \citet{2010AJ....139.1808S} and (30.2, 16.5, 15.8) \kms from \citet{2007ApJ...666.1205Z}.

Our objects spend their time in a galactic potential which will vary and lead to dynamical
evolution. A monotonic increase of velocity dispersion with age is seen for a given stellar population. We studied the kinematic age following the relationship between velocity dispersion and age given in \cite{2008gady.book.....B}:

\begin{equation}\label{age08}
    \sigma_v(\tau) = v_{10} \left( \frac{\tau + \tau_1}{10{\rm Gyr} + \tau_1} \right)^{\beta}
,\end{equation}

where $\sigma_v$ is the unweighted total velocity dispersion, and we used all six best-fit parameter sets of $v_{10}$, $\tau_1$ and $\beta$ in Table 2 of \cite{2009MNRAS.397.1286A} to provide an average age in ($\tau$). With $\sigma_v = 36.75^{+4.21}_{-4.07}$\,\kms, we find $\tau =2.97^{+0.90}_{-1.06}$\,Gyr (listed in Table \ref{tab:kine_distr}). This value is similar to the value of 3.1 Gyr for a sample of 63 late M dwarfs\citep{2009ApJ...705.1416R}, but older than 1.7 Gyr for a sample of nearby L and T dwarfs \citep{2018PASP..130f4402W} -- consistent within 3$\sigma$ for both values.

\section{Binaries in our sample
\label{sec:cpm}}

Among the sources in Table \ref{tab:BDs}, one particular group deserves special attention: objects in binary systems. We study them using the GEDR3 and Pan-STARRS DR1 \citep[here after PS1,][]{2016arXiv161205560C} data. In this subsection, we report 35 ``co-moving'' binaries. The following subsections describe how we selected the binaries and carried out our analysis using the CAMD diagram, along with our  literature search for previously reported binaries.

\subsection{Identification of the binaries
\label{sec:sample_binary}}

\begin{figure}
\centering
\includegraphics[width=89.3mm]{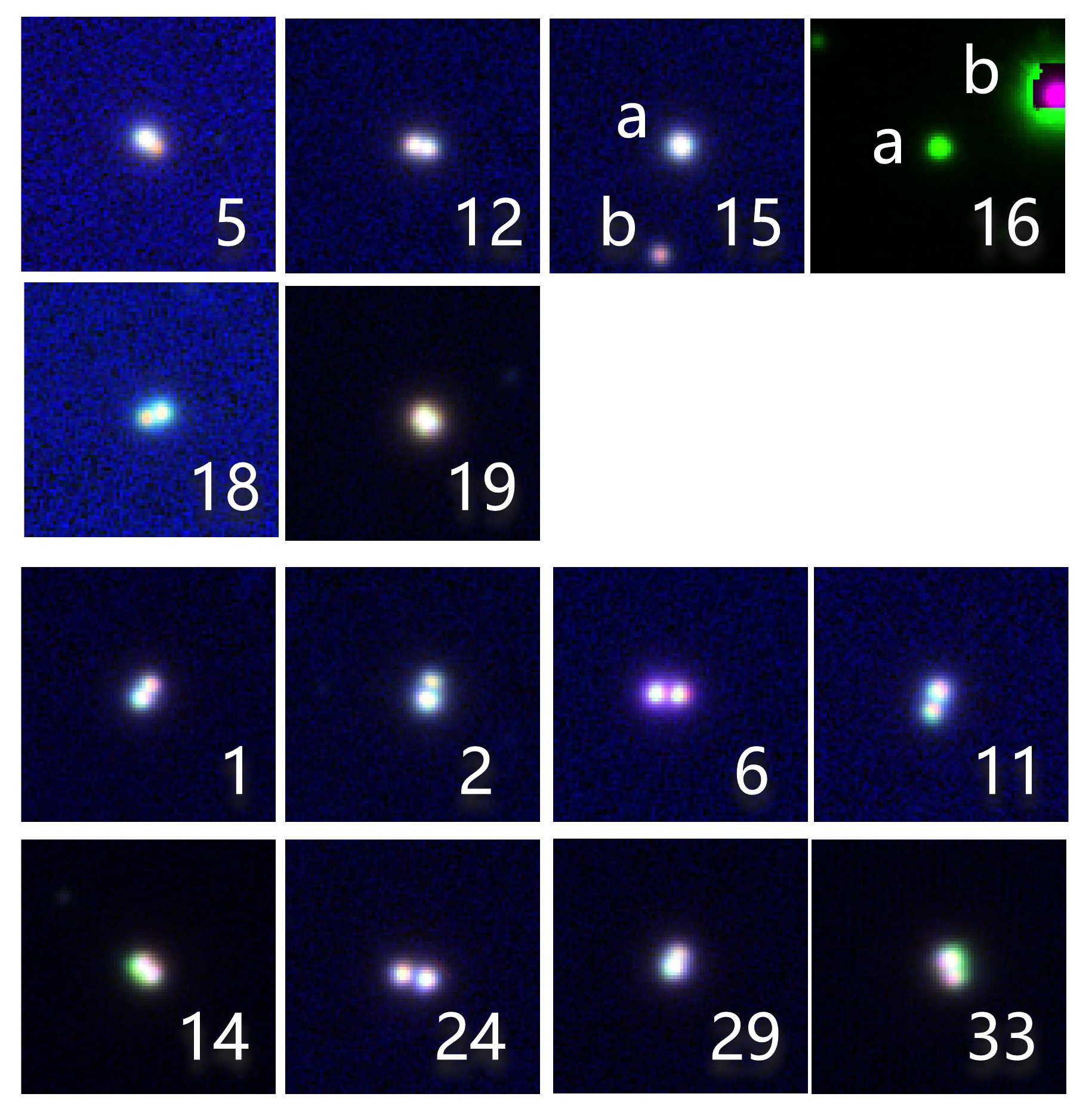}
\caption{\label{fig:cpm} PS1 y/i/g band combined images for the 14 newly discovered binaries. The first two rows of images (labeled 5, 12, 15, 16, 18, and 19) show the binaries discovered directly in this work and the last two rows of images show those found by \citet{2021A&A...649A...6G}.} With north at top south at the bottom, west at right, and east at left. Their number IDs in Table \ref{tab:cpm} are labeled in each frame. The scale is 20 arcsec for all these binary frames. The component ``$a$'' is always at the center of each image. The positions for the two components are labeled out in two frames(for those that have a wider separation).
\end{figure}

Every source in Table ~\ref{tab:BDs} has been queried in GEDR3 for possible companions within an angular radius of 16\arcsec. We then selected the systems with components comparable in proper motion and parallax; these are ``common proper motion'' binaries. We checked all images in PS1, but in Figure~\ref{fig:cpm}, we only present the PS1 images of six newly discovered binaries in this work and the eight newly discovered ones by \citet{2021A&A...649A...6G}. Table~\ref{tab:cpm} summarizes the properties of all our binaries. In this table, the first column is the identification number, with component ``$a$'' being the entry in Table~\ref{tab:BDs}, namely, having a LAMOST spectrum. Component  ``$b$'' is the co-moving object in the system, with no implication of apparent brightness or mass relative to the component ``$a$.'' Column 2 lists the GEDR3 source\textunderscore id, with columns 3 and 4 being the coordinates in J2000 epoch computed the GEDR3 epoch J2016.0; Columns 5 to 10 list the GEDR3 $G$-band magnitude, parallax, proper motion, RUWE, and the angular separation (calculated using the positions in columns 3 and 4. Column~11 lists the references.

We note that the diameter of the LAMOST fiber is 3.3 arcsec and due to fiber packing, it is not feasible to target each component in a very close binary or multiple systems. We note that in this sample, every LAMOST object in Table \ref{tab:BDs} and its ``companion'' from the GEDR3 are resolved by GEDR3. But only the LAMOST ``a'' component in the pair was targeted by LAMOST, so its spectral type is known as listed in Table ~\ref{tab:BDs}.

\subsection{Positions in the CAMD
 \label{sec:new_discovered}}

\begin{figure}
\centering
\includegraphics[width=89.3mm]{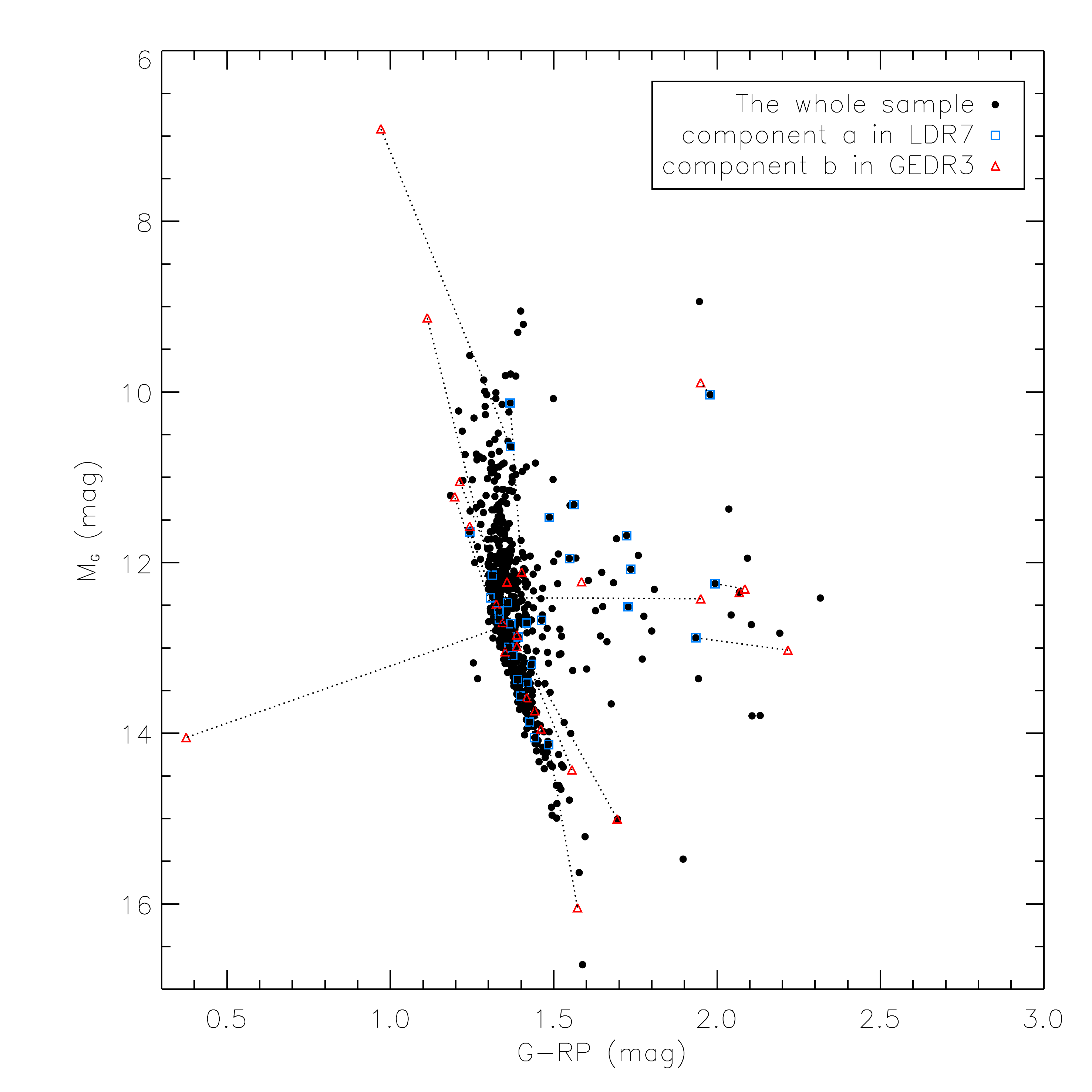}
\caption{\label{fig:cmp_hrd} CAMDin the \textit{Gaia} photometry system for binary components. The x-axis is $G-G_{RP}$ and the y axis is absolute $G$ magnitude. The legends at the top right describe the meaning of each symbol.}
\end{figure}

There are 35 binaries in total and 6 of them are newly discovered in this work, with images presented in Figure \ref{fig:cpm}. Below, we show the positions for all of these binaries in the CAMD. With the magnitudes and parallaxes from the GEDR3, we studied the binaries in the CAMD. To plot the CAMD, we require the objects have relative uncertainty on a parallax better than 20$\%$ and \textit{Gaia} $G$ and $G_{RP}$ magnitudes greater than zero.

In Figure \ref{fig:cmp_hrd}, the black solid dots indicate the whole sample. While the blue squares indicate the LAMOST ''a'' components; the red triangles indicate the GEDR3 ''b'' components. The ``a'' and ``b'' components are connected using the dotted lines: there are 21 lines. The left 14 binaries have at least one component lacking a parallax or magnitude measurements from GEDR3, leading that only one component appears on this figure without lines connecting the two components. The dotted lines are largely parallel to the main sequence, which proves that they are coeval binaries, though some lines are far from the main sequence. There is one ``b'' component at the white dwarf loci: ``26b''. The ``26ab'' is a known white dwarf-main-sequence binary reported by \citet{2020ApJS..246....4T}. There are several lines not parallel to the main-sequence, which indicate that the components with significantly different colors might have different metallicities and ages.  Besides, objects with redder colors and brighter magnitudes than the main sequence might be unresolved binaries.

\subsection{Previously known systems
\label{sec:known}}

To search for known binaries, we first looked into the WDS catalog, the successor to the Index Catalogue of Visual Double Star, maintained by the United States Naval Observatory. The catalog provides parameters for double and multiple systems. There are 16 systems in Table~\ref{tab:cpm}, already listed in the WDS catalog. Another binary catalog that we cross-matched with is the ``Catalog of ultrawide binary stars from \textit{Gaia} DR2'' \citep{2020ApJS..246....4T}, in which we found seven of our binaries. The third catalog we queried is the ``SUPERWIDE Catalog'' by \citet{2020ApJS..247...66H}, in which six of our binaries are listed. In the fourth catalog, we searched in Table 3 from \citet{2021A&A...649A...6G}, and found 19 of our binaries. In the literature, these are known as a ``binary'' that is worthy of attention for its particular properties.

System No.~21, known as $LP\,213-67$ and $LP\,213-68$, has a rich record in the literature and might prove to be a quadruple system \citep{2017ApJS..231...15D}. It was first discovered by \citet{1980PMMin..55....1L} as a pair of ultracool dwarf (M6.5 for $LP\,213-67$ and M8 for $LP\,213-68$), and later investigated by \citet{2003ApJ...587..407C} using the AO system on Gemini to resolve the LP\,213-68 \footnote{Originally in \citet{2003ApJ...587..407C},
they reported an observation of $LP\,213-67$ but \citet{2017ApJS..231...15D} found that the separation and position angle derived by \citet{2003ApJ...587..407C} is much closer with their study on $LP~213-68$ rather than the result of $LP~213-67$ by \citet{2010ApJ...711.1087K}}. \citet{2017ApJS..231...15D}, with a decade-long astrometric monitoring image from Keck Observatory, Hubble Space Telescope, and CFHT/WIRCam, suggested that $LP\,213-68$ may be a system with a main-sequence star having mass $  \sim97$ M$_{Jup}$ plus a BD $  \sim80$ M$_{Jup}$ with a $  \sim 2397$-day period. On the other hand, \citet{2010ApJ...711.1087K} tried to use the Keck\, II 10\,m telescope with the LGS AO system to derive the orbit and period for $LP\,213-67$, but failed with insufficient epoch coverage. They provide a separation of 0\farcs032 $\pm$ 0\farcs002 and position angle of  $\sim$ 126.77$\degr$ on 2006 Jun 21. A more detailed summary on $LP\,213-67$ and $LP\,213-68$ can be found in \citet{2017ApJS..231...15D}. We note that the "b" component of this binary system has a red color and brighter absolute G magnitude; furthermore, the RUWEs of the two components is greater than 1.4. The above color, absolute magnitude, and RUWE information could be indications of an unresolved binary for each individual component for other wide binary systems.


\section{Summary
\label{summary}}

LAMOST DR7 released 11,412 spectra with the SpT equal to or later than M6, with both giants and dwarfs included. We visually inspected individual spectra according to spectral features and selected 734 ultracool dwarfs by excluding giants, misclassified early-type M dwarfs, and low S/N spectra. Based on the SIMBAD database, about 82\% of the ultracool dwarfs have not been measured their spectral type before LAMOST.

We cross-matched all these 11,412 entries with the~ \textit{GEDR3} within 5 arcsec and obtained their position on the CAMD with magnitudes and parallax of \textit{Gaia}.  We found that these samples show their correct locus as cool dwarfs on both the CAMD and the $BP-RP$ versus $G-RP$ diagram.  The ultracool dwarfs are within 360 pc(83\% of our ultracool dwarf sample are within 100pc) and brighter than 19.2 in \textit{Gaia} G magnitude. The mean value of G band magnitude is around 16 mag.

We analyzed the stellar parameter measurements from LDR7. Effective temperature, surface gravity, and metallicity exhibit trends that are consistent with the literature expectation. Since these parameters are not calibrated for lacking sufficient external parameter estimation, it is important that individual spectra are checked before adopting the parameters. We did not estimate the mass for the ultracool dwarfs using stellar parameters and evolutionary tracks, but we checked the lithium line at 6708 \AA using the LAMOST low-resolution spectra. Overall, 10\% of our ultracool dwarfs show a lithium absorption line, which is an indication of the young and substellar nature of ultracool dwarfs.

With the RV from LDR7 and astrometric data from GEDR3, we studied the Galactic membership of these ultracool dwarfs. From the Galactic position, Toomre diagram, and the probability ratio of thick-disk-to-thin-disk (TD/D) and thick-disk-to-halo (TD/H), we drew the same conclusion: they are all disk objects and most of them belong to the thin disk. This is consistent with the distances and metallicity distribution of LDR7. Using 477 thin-disk objects, we studied their $UVW$ distribution. The mean and dispersion of the velocities are derived as $(U,V,W)_{\bigodot}=[14.02^{+1.20}_{-1.21},17.60^{+0.71}_{-0.71},6.19^{+1.21}_{-1.20}]$~\kms and $[\sigma_{U},\sigma_{V},\sigma_{W}]=[25.71^{+1.17}_{-1.18},18.53^{+0.89}_{-0.89},18.60^{+1.43}_{-1.37}]$. The total standard deviation in velocity is $\sigma_{v}=36.75^{+4.21}_{-4.07}$~\kms, corresponding to a kinematic age of $\tau =2.97^{+0.90}_{-1.06}$\,Gyr.

We found 35 common proper motion binaries and 6 of them are newly discovered binaries. Those with both parallax and magnitudes are shown on the CAMD: one literature source is known ultracool dwarf-white dwarf binary and its location is as expected. Furthermore, several binaries are shown to be coeval binaries and several of them appear to exhibit a peculiar color.


\begin{acknowledgements}

We thank Mr. Shih-Yun Tang for the idea of binarity study and help with \textit{Gaia} data usage. We thank Dr. Richard L. Smart for a very useful discussion on the kinematic study of nearby cool objects. This work is supported by the National Key R\&D Program of China(No. 2019YFA0405502), National Science Foundation of China (No U1931209), Chinese Space Station Telescope(CSST) pre-research projects of {\it Key Problems in Binaries} and {\it Chemical Evolution of the Milky Way and its Nearby Galaxies}. WPC acknowledges the financial support from the grant MOST 106-2112-M-008-005-MY3. HRAJ acknowledges support from a CAS PIFI Grant No. 2020VMA0033 and UK STFC grant ST/R006598/1. YXG acknowledges the National Science Foundation of China (Nos U2031137). YBL acknowledges the science research grants from the China Manned Space Project with NO.CMS-CSST-2021-A10,NO.CMS-CSST-2021-A08 and NO.CMS-csst-2021-B05.

Guoshoujing Telescope (the Large Sky Area Multi-Object Fiber Spectroscopic Telescope LAMOST) is a National Major Scientific Project built by the Chinese Academy of Sciences. Funding for the project has been provided by the National Development and Reform Commission. LAMOST is operated and managed by the National Astronomical Observatories, Chinese Academy of Sciences.

\end{acknowledgements}

\bibliographystyle{aa} 
\bibliography{ref.bib} 

\newpage
\onecolumn
\begin{flushleft}

\begin{appendix}
\section{Binaries}

\scriptsize
\begin{longtable}{lccc cccc cccc}
\caption{\label{tab:cpm} Binaries in the sample.} \\ \hline
No.& source $\textunderscore$ id& R.A. & Decl. & $G$ & $\varpi$       & $\mu_\alpha \cos\delta$ & $\mu_\delta$ & RUWE& Sepa & LAMOST& REF \\
 &   &\multicolumn{2}{c}{(deg)} &(mag)& (mas) &\multicolumn{2}{c}{(mas~yr$^{-1}$)} & &(arcsec) &(SpTy)& \\ \hline
1a&374266821724115840&14.60929&40.96502&16.563&10.562&65.447&-30.010&1.055&1.30&dM6&4\\
1b&374266821723253120&14.60905&40.96529&17.154&10.640&69.022&-29.769&0.866&- &- &4\\\hline
2a&402667207072027008&16.21190&49.46696&16.265&10.983&-72.543&-48.021&1.121&1.50&dM6&4\\
2b&402667202774679808&16.21172&49.46733&17.297&10.916&-68.486&-48.482&0.948&- &- &4\\\hline
3a&321982840504952448&19.87751&38.24850&17.648&11.075&0.015&-47.740&0.893&5.87&dM6&2;4\\
3b&321982844799985536&19.87646&38.24974&17.630&11.045&0.255&-48.071&1.011&- &- &2;4\\\hline
4a&289697678714664320&20.60659&22.07549&16.098&18.790&-21.133&-11.048&1.209&7.33&dM6&3\\
4b&289697678714664192&20.60783&22.07711&14.670&18.848&-21.140&-12.270&1.026&- &- &3\\\hline
5a&316277788200150400&24.21901&33.36533&17.688&7.123&46.367&-0.923&1.171&1.09&dM6&-\\
5b&316277788200960128&24.21876&33.36516&19.207&7.158&45.694&-1.860&1.022&- &- &-\\\hline
6a&96492285755736448&25.81853&20.82407&16.421&17.828&144.663&-18.064&1.188&1.86&dM6&4\\
6b&96492285755202688&25.81801&20.82404&16.580&18.012&146.567&-14.172&0.883&- &- &4\\\hline
7a&342530930417172224&29.08084&37.48177&15.929&22.216&203.994&-50.869&0.966&7.36&dM7&1;4 \\
7b&342530930417172096&29.07952&37.48021&15.747&22.257&205.307&-49.469&1.166&- &- &1;4\\\hline
8a&345777204139584384&29.50488&41.38132&15.137&30.660&-35.967&20.361&1.047&3.03&dM6&1\\
8b&345765453109062784&29.50516&41.38053&15.274&30.592&-34.291&23.813&1.209&- &- &1\\\hline
9a&126144258931133568&39.74078&25.22135&14.728&30.488&208.525&-51.629&1.179&2.82&dM7&1;4\\
9b&126144258931133440&39.74023&25.22191&17.014&30.398&211.976&-49.911&1.156&- &- &1;4\\\hline
10a&433763835445730944&42.96403&44.56884&14.782&-&-&-&-&8.63&dM6&1\\
10b&433763831149781376&42.96188&44.56989&15.413&32.648&145.710&-90.795&1.207&- &- &1\\\hline
11a&3262741489171756160&50.45314&-0.86917&17.287&12.074&71.038&-109.761&0.971&1.55&dM6&4\\
11b&3262741489172243328&50.45300&-0.86876&16.806&12.117&74.970&-108.469&0.986&- & -&4\\\hline
12a&216686773736734336&55.75914&32.12439&17.342&3.451&4.907&-6.991&1.033&1.24&gM9&-\\
12b&216686773735019264&55.75880&32.12433&17.324&3.262&4.859&-7.500&0.843&- &- &-\\\hline
13a&3313784735944668544&66.91222&17.30882&15.627&9.768&19.060&-34.859&1.197&14.16&gM9&3\\
13b&3313784735944668032&66.91387&17.31239&13.162&9.561&19.384&-34.017&1.111&- &- &3\\\hline
14a&3406246551375179392&71.25115&17.27227&16.194&16.229&69.710&-24.550&1.079&1.01&dM6&4\\
14b&3406246551373367168&71.25095&17.27208&16.225&16.471&70.797&-22.375&1.000&- &- &4\\\hline
15a&3390528895217011712&79.88052&14.72899&16.591&5.100&-1.850&-8.109&1.164&8.74&dM6&-\\
15b&3390528890920404736&79.88102&14.72661&18.416&5.480&-1.689&-7.719&1.007&- &- &-\\\hline
16a&3446362851590055808&82.80520&30.94254&16.556&6.561&3.024&-28.840&0.990&11.60&dM6&-\\
16b&3446362916013895552&82.80219&30.94369&12.644&7.148&5.606&-26.952&4.962&- &- &-\\\hline
17a&3349802297332186496&89.32967&17.14099&15.428&48.737&233.929&-263.654&1.044&7.26&dM6&1;2;3;4\\
17b&3349805247975323392&89.32774&17.14157&12.785&48.804&242.615&-260.695&1.241&- &- &1;2;3;4\\\hline
18a&884269552193012096&105.63858&27.68727&17.734&9.054&-34.352&-37.865&0.860&1.27&dM6&-\\
18b&884269552190985344&105.63892&27.68717&18.312&8.646&-34.871&-39.144&1.005&- &- &-\\\hline
19a&3141512047219196672&113.14261&6.11010&16.784&14.341&-50.159&-33.730&1.037&0.75&dM6&-\\
19b&3141512051515263872&113.14247&6.10994&16.613&14.031&-46.712&-33.758&2.795&- &- &-\\\hline
20a&660597997696173440&134.56281&19.76340&11.966&194.144&-767.060&-100.176&2.574&2.66&dM6&3\\
20b&660597997697274752&134.56311&19.76273&12.486&196.262&-937.133&-34.559&3.539&- &- &3\\\hline
21a&779689606794219136&161.80254&40.44548&15.159&40.432&-298.288&-33.258&1.498&18.37&dM7&1;4\\
21b&779689533779300736&161.80740&40.44702&17.023&39.475&-301.661&-33.982&2.785&- &- &1;4\\\hline
22a&3809562019331072384&162.94824&2.89455&15.647&35.030&-350.031&-56.090&1.011&3.53&dM7&1;4\\
22b&3809562019330343040&162.94878&2.89537&15.867&34.915&-342.425&-55.518&1.027&- &- &1;4\\\hline
23a&729808234575686912&163.62886&25.96367&15.645&25.827&-85.038&-114.002&1.012&2.26&dM6&2;4\\
23b&729808230280106752&163.62946&25.96346&15.800&25.828&-90.340&-117.304&0.975&- &- &2;4\\\hline
24a&3699132156679086720&181.09290&0.88691&16.980&13.642&93.048&-64.666&1.176&1.90&dM6&4\\
24b&3699132156679086848&181.09238&0.88680&16.624&13.189&93.822&-66.237&1.092&- & -&4\\\hline
25a&4015042738059517440&185.36273&30.64324&13.383&83.504&-209.779&-267.270&0.926&5.32&dM6&1;2\\
25b&4015042738059517568&185.36136&30.64379&13.436&83.463&-191.036&-253.760&1.167&- & -&1;2\\\hline
26a&4010091053083980032&186.44945&28.60449&17.268&12.288&-85.753&-259.333&1.057&17.51&dM6& 3;4\\
26b&4010090670831602432&186.44531&28.60193&18.565&12.500&-87.315&-258.321&0.879&- &- &3;4\\\hline
27a&4011353223713622528&189.36063&29.87850&16.547&20.354&-156.452&-102.041&1.000&10.98&dM6& 1;2;3;4\\
27b&4011353223713630592&189.36366&29.87879&12.594&20.291&-156.509&-98.555&1.384&- &- &1;2;3;4\\\hline
28a&1466583795164173184&198.02211&32.22596&16.053&18.687&113.194&-88.446&1.360&1.00&dM6&1;4\\
28b&1466583795162485120&198.02190&32.22578&16.073&18.623&112.014&-87.261&1.093&- &- &1;4\\\hline
29a&3664079485466748288&207.56108&1.50458&16.384&13.764&57.639&-40.418&2.049&1.01&dM6&4\\
29b&3664079519825981056&207.56099&1.50484&17.223&13.431&54.726&-41.720&0.833&- &- &4\\\hline
30a&3674243817630610304&212.46527&7.40705&15.128&35.534&-210.456&-36.320&0.704&1.38&dM6&1\\
30b&3674243817629910016&212.46545&7.40739&15.282&35.373&-222.378&-47.723&0.764&- &-& 1\\\hline
31a&1477762323724646272&214.94243&31.61928&15.442&52.662&104.455&-21.802&1.125&9.87&dM7&3\\
31b&1477762319428998400&214.94467&31.61770&12.962&52.747&99.281&-25.453&1.320&- &- &3\\\hline
32a&1267909667590644992&225.58881&25.43186&18.204&15.332&-124.062&-54.785&1.128&9.02&dM7&2 ;4\\
32b&1267909671886228224&225.58885&25.43437&20.059&15.745&-123.180&-54.482&1.081&- &- &2;4\\\hline
33a&1271532528341730176&228.43425&27.39515&15.559&14.194&-54.379&19.133&5.421&1.31&dM6&4\\
33b&1271532528339403648&228.43417&27.39480&16.390&14.219&-59.596&23.552&0.983&- &- &4\\\hline
34a&1221808523727348480&232.99323&24.75628&17.206&7.709&-71.250&30.714&0.978&3.34&dM6&1;2\\
34b&1221808519429702784&232.99400&24.75679&19.335&7.591&-71.475&30.192&1.081&- &- &1;2\\\hline
35a&2079073928612821760&298.47977&44.41504&11.908&214.574&349.363&-480.322&1.611&9.75&dM7&1\\
35b&2079074130463898624&298.47977&44.41233&11.535&-&-&-&-&-&-&1\\ \hline
\end{longtable}

\tablefoot{Column 1 the identification(ID) number of the binaries, arranged in the order R.A. rising. Column 2:  GEDR3 source$\textunderscore$ id. Columns 3-4: R.A. and Decl. in J2000(transferred from the GEDR3 epoch J2016.0). Columns 5-9: \textit{Gaia} G band magnitude, parallax, proper motion, and RUWE from GEDR3. Column 10: Eeparation of the two components calculated using the position in columns 3 and 4. Column 11: References from 1-4: \citet{2001AJ....122.3466M,2020ApJS..247...66H,2020ApJS..246....4T,2021A&A...649A...6G}. The six objects without a reference are newly discovered.}

\end{appendix}
\end{flushleft}

\twocolumn
\end{document}